\documentclass[
	      ]{jfm}

\usepackage{graphicx}
\usepackage{natbib}

\ifCUPmtlplainloaded \else
  \checkfont{eurm10}
  \iffontfound
    \IfFileExists{upmath.sty}
      {\typeout{^^JFound AMS Euler Roman fonts on the system,
                   using the 'upmath' package.^^J}%
       \usepackage{upmath}}
      {\typeout{^^JFound AMS Euler Roman fonts on the system, but you
                   dont seem to have the}%
       \typeout{'upmath' package installed. JFM.cls can take advantage
                 of these fonts,^^Jif you use 'upmath' package.^^J}%
       \providecommand\upi{\pi}%
      }
  \else
    \providecommand\upi{\pi}%
  \fi
\fi

\ifCUPmtlplainloaded \else
  \checkfont{msam10}
  \iffontfound
    \IfFileExists{amssymb.sty}
      {\typeout{^^JFound AMS Symbol fonts on the system, using the
                'amssymb' package.^^J}%
       \usepackage{amssymb}%
         \let\leq=\leqslant
         \let\geq=\geqslant
      }{}
  \fi
\fi

\ifCUPmtlplainloaded \else
  \IfFileExists{amsbsy.sty}
    {\typeout{^^JFound the 'amsbsy' package on the system, using it.^^J}%
     \usepackage{amsbsy}}
    {\providecommand\boldsymbol[1]{\mbox{\boldmath $##1$}}}
\fi

\newcommand\hz{\nobreak\mbox{$\;$Hz}}
\newcommand\khz{\nobreak\mbox{$\;$kHz}}

\providecommand\bv[1]{\boldsymbol{#1}}

\newcommand\Rey{\mbox{\textit{Re}}}  

\newsavebox{\astrutbox}
\sbox{\astrutbox}{\rule[-5pt]{0pt}{20pt}}

\newcommand\p{\ensuremath{\partial}}

\newcommand\eg{e.g.\ }
\newcommand\eqref[1]{(\ref{#1})}
\newcommand\avg[1]{\ensuremath{\left<#1\right>}}
\newcommand\avga[1]{\ensuremath{\left<#1\right>_{A(r),t}}}
\providecommand{\e}[1]{\ensuremath{\times 10^{#1}}}

\title[Torque in Taylor-Couette Flow]{Direct Numerical Simulations of Local and Global Torque in Taylor-Couette Flow up
to Re=30.000}

\author[H. J. Brauckmann and B. Eckhardt]%
{H\ls A\ls N\ls N\ls E\ls S\ns J.\ns B\ls R\ls A\ls U\ls C\ls K\ls M\ls A\ls N\ls N$^1$
\\ \and
B\ls R\ls U\ls N\ls O\ns E\ls C\ls K\ls H\ls A\ls R\ls D\ls T$^{1,2}$%
  \thanks{Email address for correspondence: bruno.eckhardt@physik.uni-marburg.de}}

\affiliation{$^1$Fachbereich Physik, Philipps-Universit\"at Marburg,
Renthof 6, D-35032 Marburg, Germany\\[\affilskip]
$^2$J M Burgerscentrum, Delft University of Technology, Mekelweg 2, 2628 CD Delft, The Netherlands}

\pubyear{2010}
\volume{650}
\pagerange{119--126}
\begin{document}

\maketitle

\begin{abstract}
The torque in turbulent Taylor-Couette flows for shear Reynolds numbers $Re_S$ up to $3\times10^4$ at various mean rotations is studied by means of direct numerical simulations for a radius ratio of $\eta=0.71$. 
Convergence of simulations is tested using three criteria of which the agreement of dissipation values estimated from the torque and from the volume dissipation rate turns out to be most demanding.
We evaluate the influence of Taylor vortex heights on the torque for a stationary outer cylinder and select a value 
of the aspect ratio of $\Gamma=2$, close to the torque maximum. 
The connection between the torque and the transverse current $J^\omega$ of azimuthal motion which can be computed from the velocity field enables us to investigate the local transport resulting in the torque. The typical spatial distribution of the individual convective and viscous contributions to the local current is analysed for a turbulent flow case.
To characterise the turbulent statistics of the transport, PDF's of local current fluctuations are compared to experimental wall shear stress measurements. PDF's of instantaneous torques reveal a fluctuation enhancement in the outer region for strong counter-rotation. Moreover, we find for simulations realising the same shear \mbox{$Re_S\geq2\times10^4$} the formation of a torque maximum for moderate counter-rotation with angular velocities \mbox{$\omega_o\approx-0.4\,\omega_i$}. In contrast, for \mbox{$Re_S\leq4\times10^3$} the torque features a maximum for a stationary outer cylinder. In addition, the effective torque scaling exponent is shown to also depend on the mean rotation state. Finally, we evaluate a close connection between boundary-layer thicknesses and the torque.
\end{abstract}

 \begin{keywords}
 \end{keywords}

\section{Introduction}
The flow between two concentric independently rotating cylinders shows, after a first bifurcation to axially periodic vortices \citep{Taylor1923}, various routes to turbulence. The presence of two easily accessible control parameters, the rotation rates
of the two cylinders, and two geometrical parameters, the ratio of the radii and the aspect ratio, the height of the cylinders relative to the gap, provide a huge parameter space with a rich phenomenology \citep{Andereck1986,Koschmieder1993,Chossat1994}. For the turbulent
situation one might expect much less of a dependence on these parameters because of the homogenising effects of
turbulent fluctuations, but already the experiments by \citet{Wendt1933}, \citet{Taylor1936} and later
studies \citep{Lathrop1992,Lathrop1992a,Lewis1999,Ravelet2010,Burin2010,Paoletti2011,VanGils2011,VanGils2011b} 
show that there is considerable structure in the data that needs to be explained. 

The measurements of torque by \citet{Wendt1933} up to Reynolds numbers $10^5$ show that for the
case of the outer cylinder at rest the dependence of the torque on the inner cylinder velocity can be described by two effective
power laws, one for high and one for low shear rates, see also \citep{Racina2006,Burin2010}. In contrast, high-precision torque measurements with a stationary outer cylinder revealed that the torque does not follow a pure power-law scaling since the local scaling exponent increases monotonically with the inner cylinder velocity \citep{Lathrop1992,Lathrop1992a,Lewis1999}. Moreover, first torque determinations also for counter-rotating cylinders and for the inner cylinder at rest showed an even more complicated dependence of the torque scaling exponent on the applied shear \citep{Ravelet2010}. In addition to the scaling with the shear rate, recent experiments for Reynolds numbers up to a few $10^6$ revealed that, when the mean shear rate is kept constant, the maximum in torque is reached 
for counter-rotating cylinders \citep{Paoletti2011,VanGils2011,VanGils2011b}. 

These high Reynolds number experiments were complemented in narrow Reynolds number regimes with direct numerical simulations (DNS) of turbulent Taylor-Couette: Flow characteristics were analysed for the outer cylinder at rest \citep{Bilson2007,Dong2007,Pirro2008}, for counter-rotating cylinders \citep{Dong2008} and in the case of spiral turbulence \citep{Meseguer2009b,Dong2009,Dong2011}. The highest Reynolds number ($\sim10^4$) was achieved by \citet{Pirro2008}, and it touches the Reynolds number range studied experimentally, 
but they do not analyse the turbulent transport and its contributions to the torque. 
Our aim here is to present direct numerical simulations of Taylor-Couette flow up to $\Rey=3\times10^4$ and to show that 
with these simulations one can explore the cross-over regime from the low, vortex dominated  $\Rey$ regime to the 
more fully turbulent regime accessible in experiments. We will be guided by previous numerical simulations of 
\citet{Coughlin1996} and \citet{Dong2007,Dong2008} and the experimental findings of 
\citet{Paoletti2011} and \citet{VanGils2011}.

The outline of the paper is as follows. In section \ref{sec:numerics} we define the problem, describe the numerical method
and explain the convergence tests we applied. The dependence of the torque on the domain height is analysed in section
\ref{sec:vortex_size}. The local contributions to the torque near the inner and outer cylinder and the variations
with the radius are discussed in section \ref{sec:av_current}. This section also includes a discussion of the fluctuation 
statistics measured by  \citet{Lewis1999}. In section \ref{sec:rot_depend} we discuss the dependence of the torque
on both Reynolds numbers and discuss the relation to recent experiments by \citet{VanGils2011,Paoletti2011,Huisman2012}. 
Finally, we discuss the determination of a boundary-layer thickness that is compatible with the definition of the
current proposed in \cite{Eckhardt2007} in  section \ref{sec:BLrelation}. We conclude with a brief summary and outlook.

\section{Problem setting and numerical methods}\label{sec:numerics}
\subsection{Definitions}
We investigate the incompressible flow between two concentric cylinders by means of direct numerical simulations. The geometry of the Taylor-Couette (TC) system is characterised by the radius ratio $\eta=r_i/r_o$, where $r_i$ and $r_o$ denote the radius of the inner and outer cylinder, respectively. Here, we 
study the case $\eta=0.71$ which is close to values chosen in recent experimental torque studies \citep{Lathrop1992,Lathrop1992a,Lewis1999,VanGils2011,Paoletti2011}. The simulations are performed in an axially periodic domain and therefore do not include the effects of top and bottom lids present in TC experiments. Unless otherwise specified, the aspect ratio $\Gamma=L_z/d=2$ is chosen, with $L_z$ the height of the cylinders and $d$ the gap width, $d=r_o-r_i$. 
This domain can support one Taylor vortex pair when the outer cylinder is at rest. 
In the figures we employ the rescaled radial coordinate $y=(r-r_i)/d$ which ranges between $0$ for the inner cylinder
and $1$ for the outer cylinder.

For the measure of the shear between the cylinders we use the definition of a shear Reynolds number $\Rey_S$ 
given in \citet{Dubrulle2005},
\begin{equation}
 \Rey_S=\frac{2}{1+\eta}\left|\eta \Rey_o-\Rey_i\right|,
\end{equation}
with $\Rey_i=dr_i\omega_i/\nu$ and $\Rey_o=dr_o\omega_o/\nu$, where $\omega_i$ and $\omega_o$ denote the angular velocity of the inner and outer cylinder and $\nu$ the kinematic viscosity of the fluid. We measure the mean (or global) rotation of the system by the ratio of angular velocities \citep{VanGils2011}
\begin{equation}
 a= -\omega_o/\omega_i .
\end{equation}
Note that for $\Rey_o=0$ the shear Reynolds number $\Rey_S=2/(1+\eta)\Rey_i$ is slightly larger than $\Rey_i$. 

To connect to the analysis in \citet{Eckhardt2007}, we consider the local current of angular velocity
\begin{equation}
  j^\omega(r,\varphi,z,t) = r^3 (u_r \omega - \nu \partial_r \omega) ,
 \label{eq:loc-current}
\end{equation}
where $u_r$ and $u_\varphi$ denote the radial and azimuthal velocity components and $\omega=u_\varphi/r$ is the angular velocity. 
Its average over any cylindrical surface concentric with the bounding cylinders gives the
mean angular velocity current 
\begin{equation}
 J^\omega = \avga{j^\omega} = r^3\left(\avga{u_r \omega}-\nu\p_r\avga{\omega} \right),
 \label{eq:current}
\end{equation}
which is related to the dimensionless torque $G=T/(2\upi L_z\rho_f\nu^2)=\nu^{-2}J^\omega$, 
where $T$ is the torque needed to drive the cylinders and $\rho_f$ is the density of the fluid.  
Here $\avga{\cdots}$ stands for averages over a cylindrical surface of radius $r$ within $r_i\leq r \leq r_o$ and over time. 
As the current is independent of $r$, any $r_i\leq r \leq r_o$ can be used to 
calculate it, but the rate of convergence varies with $r$ (see below). 

A rescaled value of the angular velocity current which is analogous to the Nusselt number describing the heat current in 
thermal convection \citep{Eckhardt2007a}
is obtained by dividing  $J^\omega$ by its laminar value for the circular Couette flow,
 $J^\omega_{lam}=2\nu r_i^2 r_o^2(\omega_i-\omega_o)/(r_o^2-r_i^2)$,
\begin{equation}
 Nu_\omega = J^\omega / J^\omega_{lam} =G/G_{lam} .
\end{equation}
The resulting quantity $Nu_\omega$ is the \textit{quasi-Nusselt number} and gives the torque in units
of the laminar value. 

In the following we will study the contributions of the different terms in (\ref{eq:current}) to the local and 
global values, the fluctuations of the local currents and the variation of the global torque with Reynolds number
and rotation ratios for shear Reynolds numbers up to $3 \times 10^4$.

\subsection{Numerical considerations}

The numerical code used for all subsequent computations was developed by \citet{Meseguer2007} and employs a solenoidal spectral basis to accomplish a Petrov-Galerkin scheme, \eg \citet{Moser1983,Meseguer2003}. For the computations all variables and fields are rendered dimensionless with the gap width $d$ and the viscous time $d^2/\nu$ as characteristic scales for space and time. In the numerical scheme the velocity field $\bv{u}=\bv{u_b}+\bv{v}$ is decomposed into its laminar contribution $\bv{u_b}$ and a deviation $\bv{v}$. The latter is expanded in a Fourier-Chebyshev vector field basis $\bv{\Phi}_{lnm}$,
\begin{equation}
 \bv{v}(r,\varphi,z,t)=\sum_{l=-L}^L \sum_{n=-N}^N \sum_{m=0}^M a_{lnm}(t)\, \bv{\Phi}_{lnm}, \qquad \bv{\Phi}_{lnm}=e^{i(l k_0 z + n \varphi)}\, \bv{v}_{lnm}(r),
 \label{eq:expansion}
\end{equation}
with $k_0=2\upi/\Gamma$ and spectral coefficients $a_{lnm}$. The basis functions $\bv{\Phi}_{lnm}$ are chosen to satisfy the incompressibility condition $\bv{\nabla\cdot\Phi}_{lnm}=0$ and boundary condition $\bv{v}_{lnm}(r_i)=\bv{v}_{lnm}(r_o)=0$ and the vector functions $\bv{v}_{lnm}$ are constructed from Chebyshev polynomials. The highest order of modes employed in axial, azimuthal, and radial direction is given by $L$, $N$, and $M$, respectively. Introducing the expansion \eqref{eq:expansion} to the Navier-Stokes equation and projecting it over a similar set of test basis fields leads to an ordinary differential equation for the coefficients $a_{lnm}$, which is solved using a 4th-order semi-implicit integration scheme.

The circumference of the outer cylinder in units of the gap is given by $2\pi r_0/d=2 \pi /(1-\eta)$. Already for $\eta=0.71$
this is larger than $21$, so that a computational domain that goes around the cylinder is very elongated. For some of the 
highest Reynolds number simulations we took advantage of the fact that also correlations in azimuthal direction
decay and restricted the flow to periodically continued azimuthal domains of size $2\pi/n_{sym}$ with $n_{sym}$ up to 9. 
In the spectral expansion \eqref{eq:expansion} this is achieved by substituting $n= n_{sym}n'$ and summing over $n'$ instead of $n$. Then all functions are $2\pi/n_{sym}$-periodic in the azimuthal direction and repeat $n_{sym}$-times to fill the complete circumference. In the turbulent case, the azimuthal velocity correlations decay over a short length and therefore only an azimuthally shorter sub-domain is required to capture the essential flow dynamics. A 
reference calculation for the case of the outer cylinder at rest showed that the difference in the torque was less than $1 \%$. 
Therefore, a suitable choice of $n_{sym}>1$ substantially decreases the computational costs and gives access to the
highest Reynolds numbers.

\subsection{Convergence criteria}
Convergence of the calculations is tested using three criteria. 

The first one exploits the radial independence of the 
current definition (\ref{eq:current}) and compares the values obtained for the inner and outer cylinders. A calculation
is considered converged when the two values agree within their fluctuations, see also \citet{Marcus1984,Dong2007}.

A second convergence criterion requires the coefficients in the Fourier-Chebyshev expansion \eqref{eq:expansion} to be sufficiently small.
In order to define a measure of the relative drop of the coefficients in different physical directions, we normalise the 
maximal amplitude of the highest mode in each direction by the globally strongest 
mode \mbox{$a_{\mathrm{max}}=\max_{l,n,m} |a_{lnm}|$}, i.e.
\begin{subeqnarray}
 \tilde{a}_L &=& \max_{n,m}\{|a_{Lnm}|,|a_{-Lnm}|\}/a_{\mathrm{max}}, \nonumber\\
 \gdef\thesubequation{\theequation} 
 \tilde{a}_N &=& \max_{l,m}\{|a_{lNm}|,|a_{l-Nm}|\}/a_{\mathrm{max}}, \\
 \tilde{a}_M &=& \max_{l,n}|a_{lnM}|/a_{\mathrm{max}}. \nonumber
 \label{eq:spectra}
\end{subeqnarray}
By this definition the 
amplitudes $\tilde{a}_L$, $\tilde{a}_N$ and $\tilde{a}_M$ provide a measure for the dynamical amplitude
range covered in the corresponding expansion. Additionally, they enable a comparison of the quality of approximation in the three directions. For a properly resolved simulation, the  values of $\tilde{a}_L$, $\tilde{a}_N$ and $\tilde{a}_M$ should be of 
comparable magnitude and sufficiently small.

A third criterion exploits energy balance: The energy injected by the torque and the driving of the cylinders at constant velocity 
must equal, in the statistically stationary state and as an average in time, 
the volume energy dissipation rate $\varepsilon$.
As shown in \citet{Eckhardt2007} this gives a relation between
$\tilde{\varepsilon}$ and the rescaled torque $Nu_\omega$,
\begin{equation}
 \tilde{\varepsilon}=\Rey_S^2 Nu_\omega \quad \mbox{with\ } \quad \tilde{\varepsilon}=d^4\nu^{-2}\left<(\p_iu_j+\p_ju_i)^2\right>_{V,t} .
 \label{eq:e_balance}
\end{equation}
Note that while \eqref{eq:e_balance} follows as an exact relation from the full Navier-Stokes equation, it is not satisfied in under-resolved numerical simulations and can therefore serve as another convergence requirement. It was previously employed by \citet{Marcus1984} as a test of numerical accuracy and its meaning for Rayleigh-B\'{e}nard convection is 
discussed in \citet{Stevens2010}.

\begin{figure}
  \centerline{\includegraphics{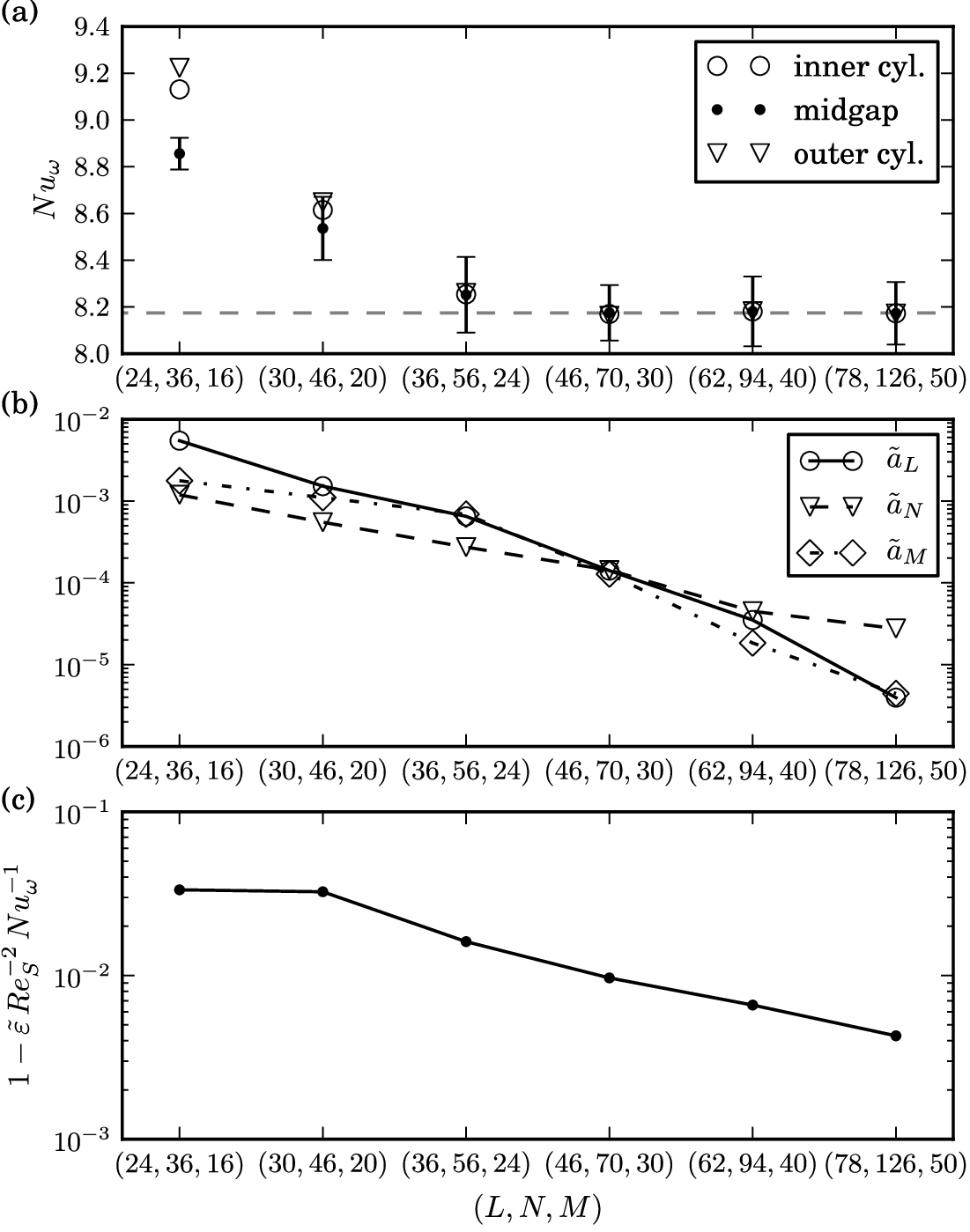}}
  \caption{Convergence analysis for a simulation with $\Rey_i=5000$ and stationary outer cylinder assessed by (a) the torque computed at three radial locations, (b) the amplitudes of the highest spectral modes and (c) the deviation from the energy balance \eqref{eq:e_balance}. In (a) the error bars for $Nu_\omega(r_a)$ denote the confidence interval due to temporal fluctuations. At the cylinder walls these confidence intervals are approximately of symbol size and are left out for this reason. The torques computed at different radii coincide for the highest four resolutions in (a).}
  \label{fig:convergence}
\end{figure}

As a first test, we consider a simulation of turbulent Taylor vortex flow for $\Rey_i=5000$ and stationary outer cylinder. 
We start with a basic resolution of $(L,N,M)=(46,70,30)$ that shows a comparable resolution in all spectral directions (cf. figure \ref{fig:convergence}(b)) as a reference and scale the number of modes up and 
down for different resolutions. Figure \ref{fig:convergence} shows the results for the three convergence criteria mentioned before. First, one observes that the quasi-Nusselt number is overestimated by under-resolved simulations, which was also noticed for Nusselt numbers in the simulation of Rayleigh-B\'{e}nard convection \citep{Stevens2010}. The three highest resolutions result in approximately the same torque which is taken as an estimate of the asymptotic 
value and indicated by the dashed grey line in figure \ref{fig:convergence}(a).
Torques calculated at the inner and outer cylinder overlap
for a resolution of $(L,N,M)=(36,56,24)$, but since the approach to the asymptotic values is monotonic from above, this agreement
is misleading and a sufficient approximation to the asymptotic value is only reached with the next step in resolution.

For resolutions above $(L,N,M)=(46,70,30)$, where the torque values does not change anymore, the amplitude of the highest mode in each spectral direction drops to $10^{-4}$, see figure \ref{fig:convergence}(b). We therefore require that a range in amplitudes of 
at least $10^4$ is covered by the spectral modes.

The third criterion on the relation between torque and energy dissipation turns out to be the most demanding: 
for the resolution $(L,N,M)=(46,70,30)$ it is only satisfied to within $1\%$ and the error falls
of rather slowly with increasing resolution. 

An analogous analysis of the convergence criteria carried out for the case of counter-rotating cylinders ($\Rey_i=-\Rey_o=2500$) 
resulted in similar findings. Following these convergence characteristics, all subsequent simulations conform with three convergence criteria: agreement of torque measurements at the inner and outer cylinder to a relative deviation of $5\times10^{-3}$, resolution of all spectral directions with relative amplitudes of the highest modes $\sim10^{-4}$ and fulfilment of the energy balance \eqref{eq:e_balance} to within $10^{-2}$.

\section{Effect of vortex sizes on the torque}\label{sec:vortex_size}
Computation of axis-symmetric Taylor vortex flow \citep{Riecke1986} show that the torque depends on the axial size of Taylor vortices. Moreover, not only the actual torque but also the scaling with Reynolds number varies for different sizes of turbulent Taylor vortices as previously measured (see figure 2 of 
\citep{Lewis1999}). Since we keep the aspect ratio small so as to be able to reach higher Reynolds numbers, we 
studied the aspect ratio dependence for the radius ratio $\eta=0.71$, once for the case of Taylor vortex flow (TVF) and then 
for turbulent Taylor vortices (TTVF). The procedure used in both cases is the following: Simulations at the same $\Rey_i$ were started for various values of the aspect ratio $\Gamma$. Then the states that formed were characterised by the number of vortices $n_v$ identified from flow visualisation. We continued by increasing or decreasing $\Gamma$ in small steps, with the flow field of the previous $\Gamma$-simulation as initial condition. The step size in terms of the axial wavelength of a Taylor vortex pair $\lambda_z = 2\,\Gamma/n_v$ did not exceed $0.1$.

For the first case of TVF, the Reynolds number $\Rey_i=150$ was chosen for two reasons: It corresponds to $\Rey_i/\Rey_c=1.85$, where $\Rey_c$ denotes the critical Reynolds number for TVF, and is thus considerably beyond the bifurcation point. This implies that a broad band of Eckhaus stable wave numbers is available and each wave number corresponds to a stable Taylor vortex state of the corresponding axial wavelength \citet{Riecke1986}. On the other hand it is below the secondary bifurcation to wavy vortex flow, which occurs for $\Rey_i/\Rey_c > 2.0$ according to a numerical analysis for $\eta=0.75$ \citep{Jones1985}. The chosen $\Rey_i=150$ is therefore in between the first and second bifurcation point and allows for the study of vortices of various sizes.
\begin{figure}
  \centerline{\includegraphics{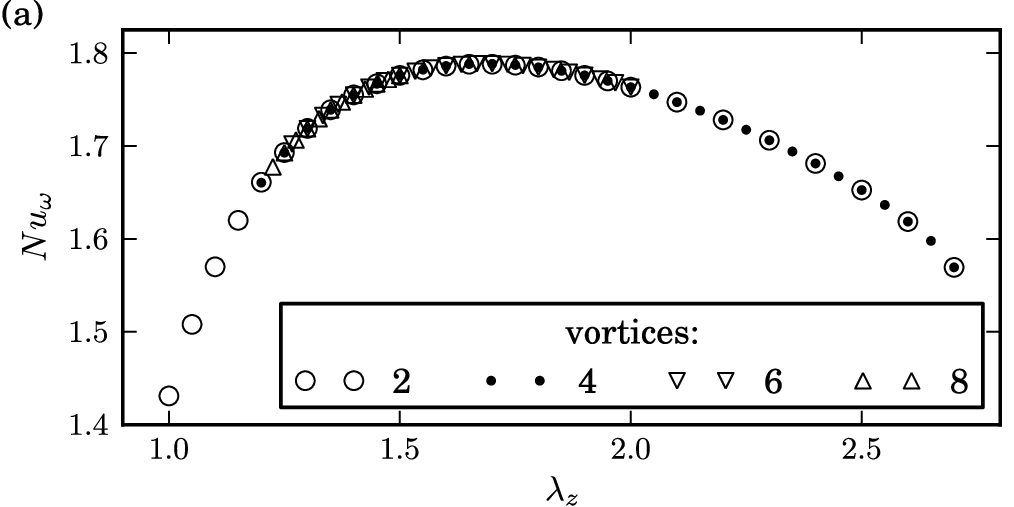}}
  \centerline{\includegraphics{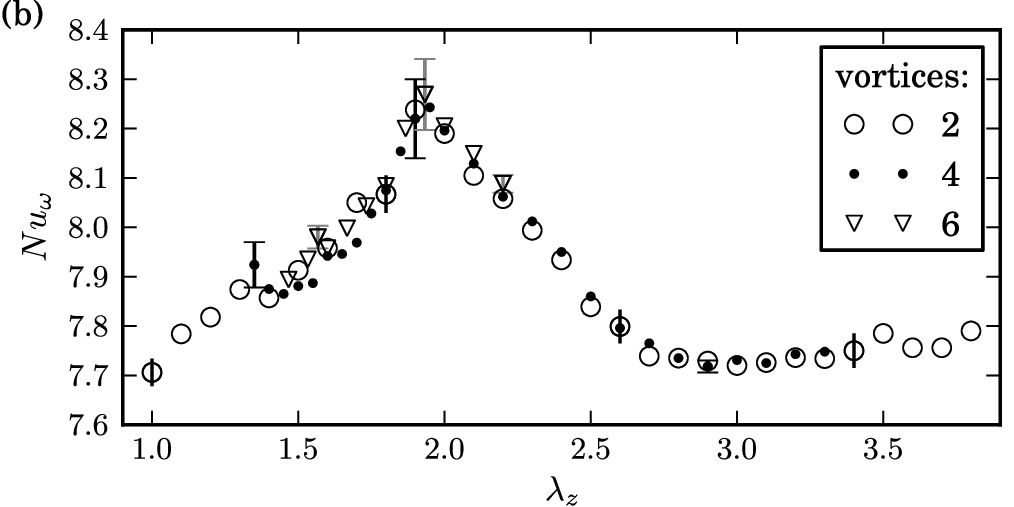}}
  \caption{The rescaled torque in dependence on the axial wavelength of vortices $\lambda_z$ (a) for the case of Taylor vortex flow ($\Rey_i=150$) and (b) turbulent Taylor vortex flow ($\Rey_i=5000$), both with stationary outer cylinder. The torques were acquired from states with various numbers of vortices. While the error bars in (b) exemplify statistical uncertainties due to fluctuations, $Nu_\omega$ was constant in (a) and no error bars are shown. }
  \label{fig:vortex-size}
\end{figure}

As a first step, states with 2, 4 and 6 vortices of the same axial wavelength $\lambda_z=2$ were simulated in aspect ratios $\Gamma=2$, $4$ and $6$, respectively. The first two states were continued by increasing $\Gamma$. This resulted in bifurcations going from 2 to 4 and from 4 to 8 vortices. Note that the latter bifurcation skipped the 6-vortex state and led to an 8-vortex state at $\Gamma=6$. These (now four different) states were followed down in $\Gamma$, and showed transitions from 8 to 4, from 6 to 4 and from 4 to 2 vortices. 
The 2-vortex state was continued to the lowest realised aspect ratio of $\Gamma=1$. 

The corresponding rescaled torque values are shown in figure \ref{fig:vortex-size}(a). Plotted against the axial wavelength $\lambda_z$ of vortices, the torque values of the different states collapse, as expected from the axial periodicity of the DNS code. $Nu_\omega$ in dependence on the vortex size exhibits a broad smooth maximum. The transport of angular velocity is maximised for $\lambda_z=1.68$, which is smaller than the critical wavelength $\lambda_c=2.0$ at the bifurcation to TVF \citep{Recktenwald1993}.

The same kind of computations for turbulent Taylor vortex flow at $\Rey_i=5000$ (with $n_{sym}=6$ to make the simulations feasible) led to similar results, see figure \ref{fig:vortex-size}(b). This time, we observed transitions between $2$ and $4$ vortices 
and from $6$ to $4$ vortices. However, a transition from $4$ to $6$ vortices did not occur within the investigated interval up to $\Gamma\leq6.6$. Again, the $Nu_\omega$ values 
collapse reasonably well in terms of $\lambda_z$. The torque is maximised for $\lambda_z=1.93$ which means an increase compared to the previous low-$\Rey_i$ value. Furthermore, the functional shape of the torque is markedly different, with a more pointed maximum, but smaller relative variations overall. 

\section{The angular velocity current}\label{sec:av_current}
The global torque is obtained as the spatial average $J^\omega$ over a local quantity, the local current 
$j^\omega(r,\varphi,z,t)$ (\ref{eq:loc-current}). The spatial and temporal fluctuations of the local quantity are influenced
by turbulent flow-structures and show different characteristics near the walls and in the middle of the gap. 
However, the averages agree as required by the radial independence of the mean $J^\omega$. 
We here analyse local properties of $j^{\omega}$
for a turbulent Taylor vortex flow at $\Rey_S=19737$ and stationary outer cylinder.

\begin{figure}
  \centerline{\includegraphics{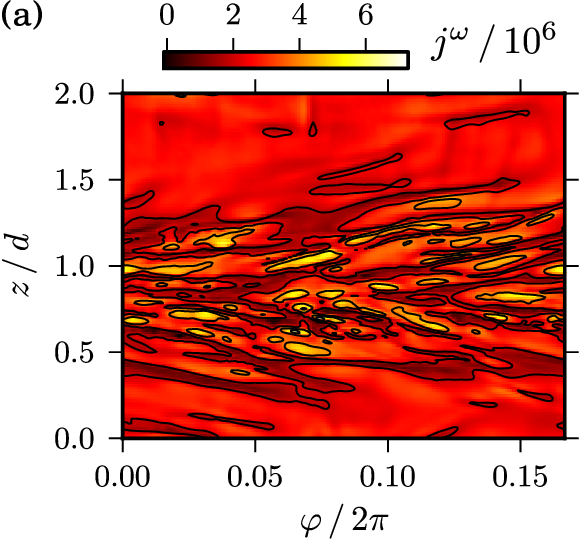}
              \hfill\includegraphics{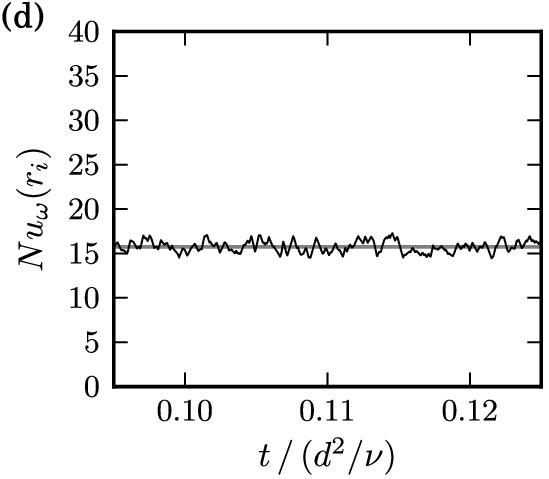}}
  \centerline{\includegraphics{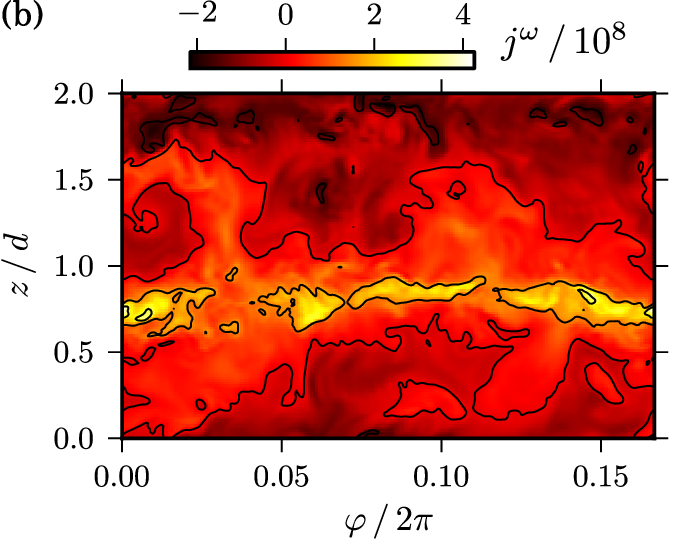}
	      \hfill\includegraphics{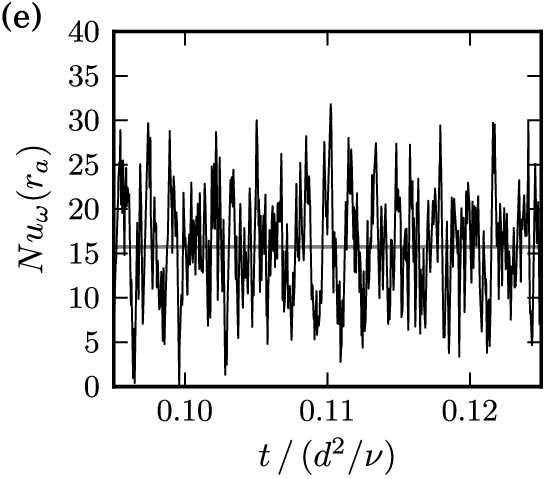}}
  \centerline{\includegraphics{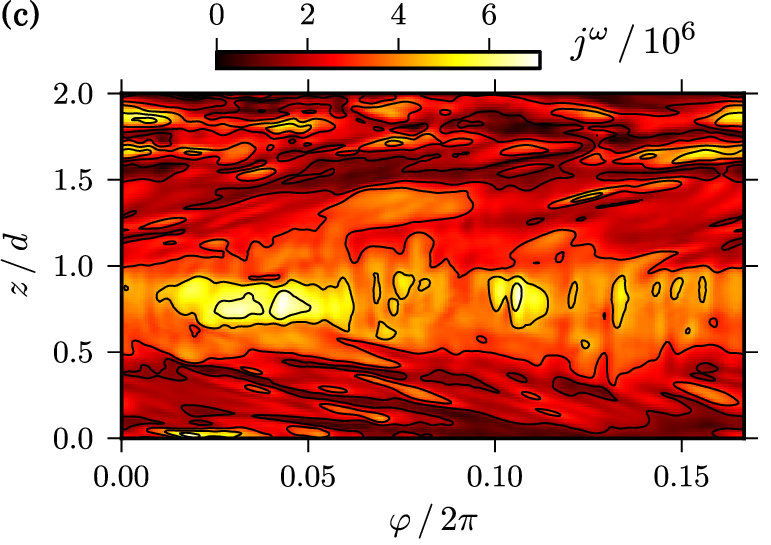}
	      \hfill\includegraphics{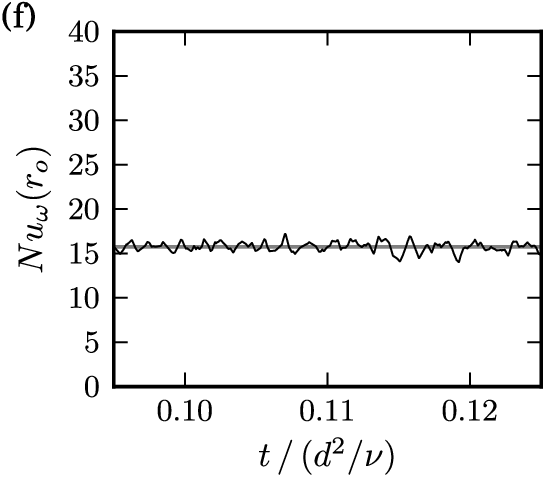}}
  \caption{Spatial and temporal fluctuations of the current $j^\omega$ resulting in the torque for $\Rey_S=19737$ and $a=0$. Cylindrical cross sections of $j^\omega$ at (a) the inner cylinder, (b) at midgap and (c) at the outer cylinder are shown and simulated for one sixth of the annulus. The abscissa always covers an angular range of $(0,2\pi/6)$ but is scaled so that the lengths reflects that the increase of the cylinder circumference with radius. The corresponding temporal fluctuations for the averaged current are shown 
  as Nusselt numbers in (d), (e) and (f). Note that the scale in (b) is about a factor $100$ larger than in (a) or (c). 
  The mean current is $J^\omega=2.62\e{6}$, the mean Nusselt number $Nu_\omega=15.7$ marked by a horizontal line in (d), (e) and (f).}
  \label{fig:current}
\end{figure}

The spatial variations of the local current $j^\omega$ adjacent to the cylinder walls and on a surface at midgap are shown in figure \ref{fig:current}(a)-(c) for a flow field of turbulent Taylor vortex flow. Note that the instantaneous torque results from an average of $j^\omega$ over such cylindrical surfaces. Comparing the section at midgap (figure \ref{fig:current}(b)) with the arrows in figure \ref{fig:contributions} reveals that the horizontal lines of low and high current $j^\omega$ correspond to the inflow and outflow regions, respectively. The spatial $j^\omega$ fluctuations at midgap are dominated by these line structures and result in extreme events that are two orders of magnitude stronger than at the cylinder walls and than the mean $J^\omega=2.62\e{6}$. 
Such extreme fluctuation amplitudes have also been seen experimentally, where it has been noted that 
the local current can exceed the mean by a factor of more than $300$ \citep{Huisman2012}.

At the origin of the outflow boundary a herringbone-like pattern of low current regions can be seen 
(figure \ref{fig:current}(a)). This pattern corresponds to the high-velocity streaks previously reported for $\eta=0.5$ in \cite{Dong2007}: High azimuthal velocity at some distance from the wall leads to a small radial derivative of $\omega$ at the inner cylinder surface. Between these herringbone stripes one can find blobs of extremely high $j^\omega$ current originating from a local steepening of the radial angular velocity gradient.

Similarly, at the outer cylinder (figure \ref{fig:current}(c)), one recovers V-shaped stripes at the origin of the inflow boundary with high $j^\omega$ blobs in between. 
Note that the $V$ is split into one part near the top and one near the bottom boundary in $z$, but that they are 
connected because of the periodic boundary condition.
 Here these stripes coincide with low-speed streaks and both stripes and blobs are axially somewhat broader than at the inner cylinder. In addition, a band of high $j^\omega$ current can be observed at the outer wall where the strong axially localised outflow ends and causes a steepening of the $\omega$-profile. Interestingly, the counterpart at the inner cylinder wall seems to be absent, which may be due to a weaker inflow.

Averaging the local transport of angular velocity over the corresponding cylindrical cross-sections results in the instantaneous torque which still exhibits fluctuations as shown in the time-series figure \ref{fig:current}(d)-(f). Again the fluctuations in the middle 
of the gap are largest. The mean torque, included as a grey line in the time-series, is obtained by additionally performing the time average for the torque signal at each radial location independently. We observe the same mean torque value despite of the different fluctuation behaviours.

\subsection{Contributions to the current}
The two contributions of convective and molecular transport in the angular velocity current \eqref{eq:current} can be further decomposed when separating the velocity field $\mathbf{u}$ into the laminar base flow $\mathbf{u}_b$ and deviating field $\mathbf{v}$,
\begin{equation}
\mathbf{u}=\mathbf{u}_b+\mathbf{v}= v_r\,\mathbf{e}_r + r\left(\Omega_b+\omega_v\right)\mathbf{e}_\varphi 
+ v_z\,\mathbf{e}_z \quad \mathrm{with} \quad \Omega_b = A+B/r^2\;.
\end{equation}
With this decomposition \eqref{eq:loc-current} becomes
\begin{equation}
 j^\omega = \underbrace{r^3(v_r\,\omega_v)}_{(1)} 
  + \underbrace{r^3 (v_r\Omega_b)}_{(2)}
  - \underbrace{\nu\,r^3 \partial_r \omega_v}_{(3)} 
  + \underbrace{2\,\nu\,B}_{(4)} \;;
  \label{eq:Jw_decomp}
\end{equation}
now with four contributions: (1) the turbulent convective transport results in a Reynolds stress-like contribution; (2) the radial transport of the laminar profile may be non-zero locally, but vanishes upon averaging due to the incompressibility condition which 
implies $\avg{v_r}_{A(r)}=0$; (3) the viscous derivative of the deviating angular velocity; and (4) the laminar term which contributes 
a constant. This motivates the definition of three fields, where the last 
two terms in (\ref{eq:Jw_decomp}) are combined,  see table \ref{tab:jwcontrib}.
\begin{table}
\centering
  \begin{tabular}{ll}
    Reynolds stress-like current & $j^\omega_{rs}=r^3v_r\,\omega_v$ \\
    convective current of the laminar profile & $j^\omega_c=r^3v_r\,\Omega_b$ \\
    viscous derivative (molecular current) & $j^\omega_{vd}=-\nu\left(r^3\partial_r\omega_v-2\,B\right)$ \\
    total $\omega$-current & $j^\omega=j^\omega_{rs} + j^\omega_{c} + j^\omega_{vd}$ \\
  \end{tabular}
  \caption{Local currents (scalar fields) contributing to $j^\omega$. The names reflect their 
           physical meanings.}
  \label{tab:jwcontrib}
\end{table}
As mentioned, the second term in \eqref{eq:Jw_decomp} contributes to the local fluctuations of the $\omega$-current, but has no impact on the instantaneous net transport through a surface $A(r)$ and, thus, does not contribute to the torque. But 
it does have a considerable impact on the local fluctuations, as we will see in the flow visualisations in the following.

\begin{figure}
  \centerline{\includegraphics{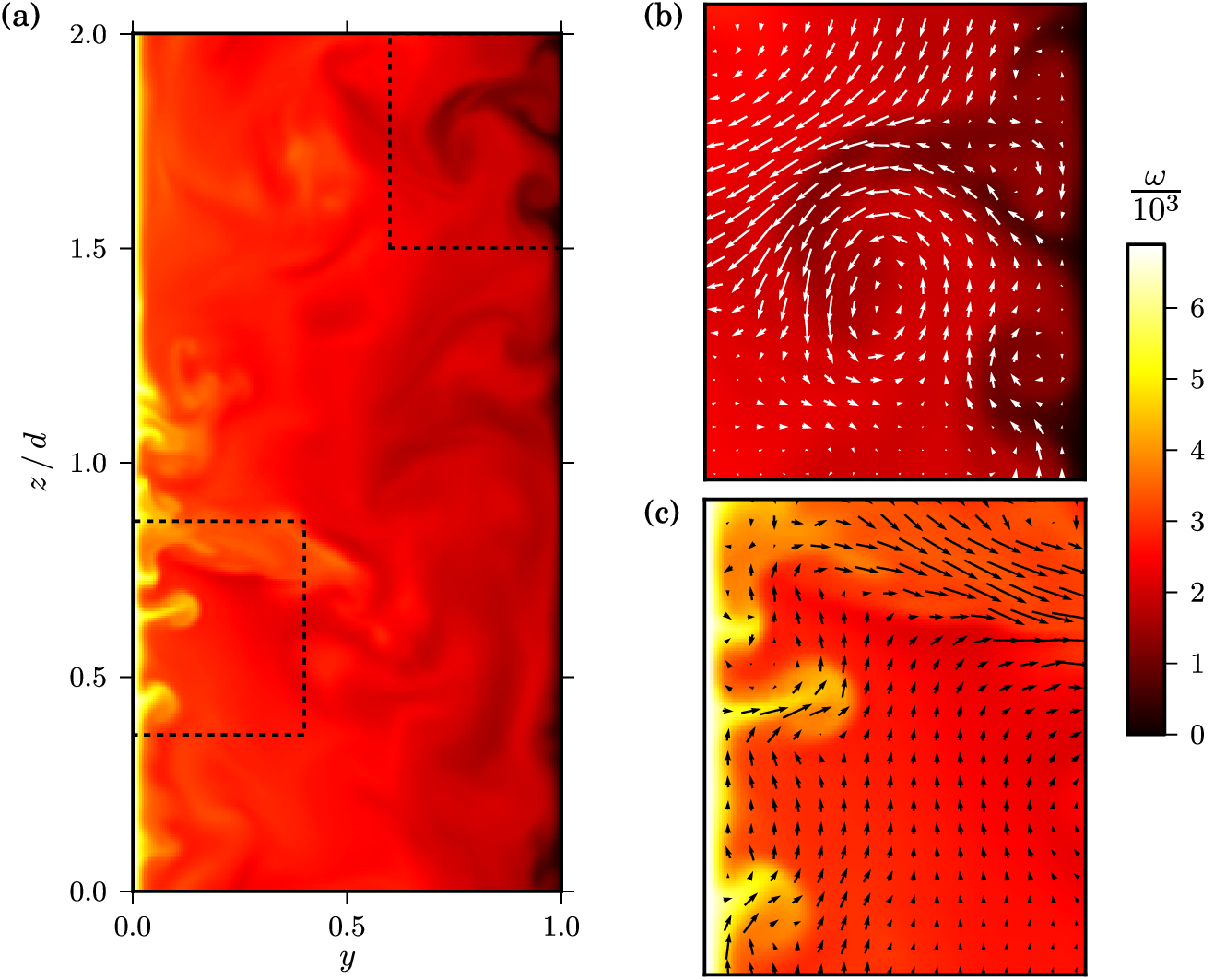}}
  \caption{Cross section of instantaneous angular velocity at a fixed angle for $\Rey_S=19737$ and $a=0$ with $y=(r-r_i)/d$. (b) and (c) are details of the marked regions in (a) at the outer and inner cylinder wall, respectively, with arrows indicating the in-plane velocity components.}
  \label{fig:flow-viz}
\end{figure}

\begin{figure}
  \centerline{\includegraphics{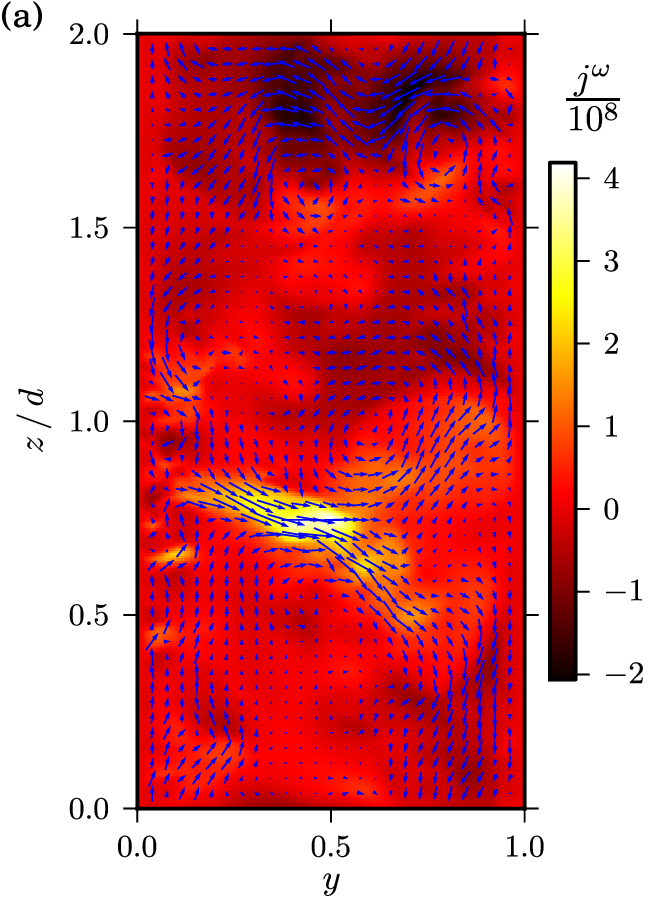}\includegraphics{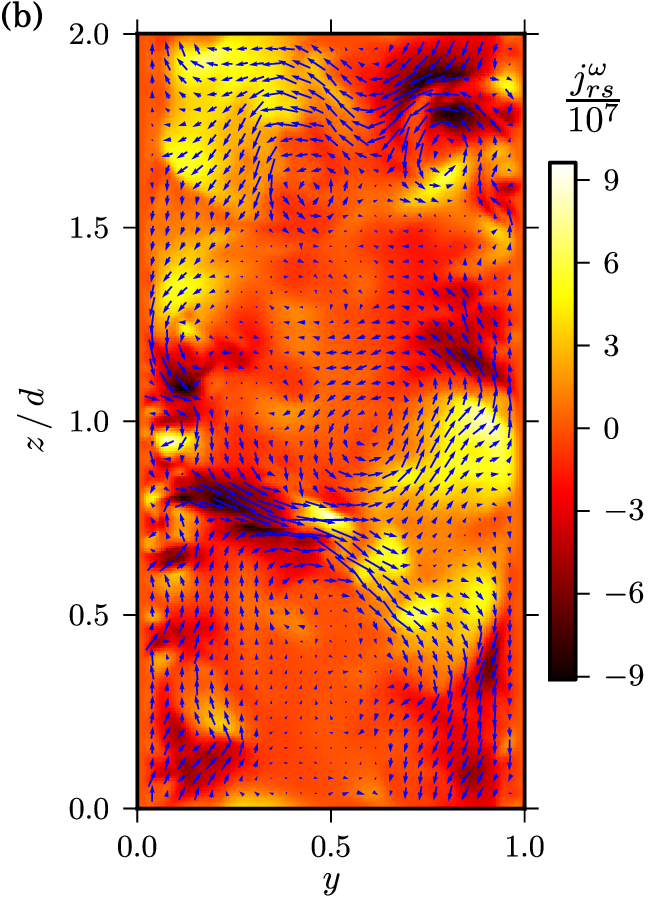}}
  \centerline{\includegraphics{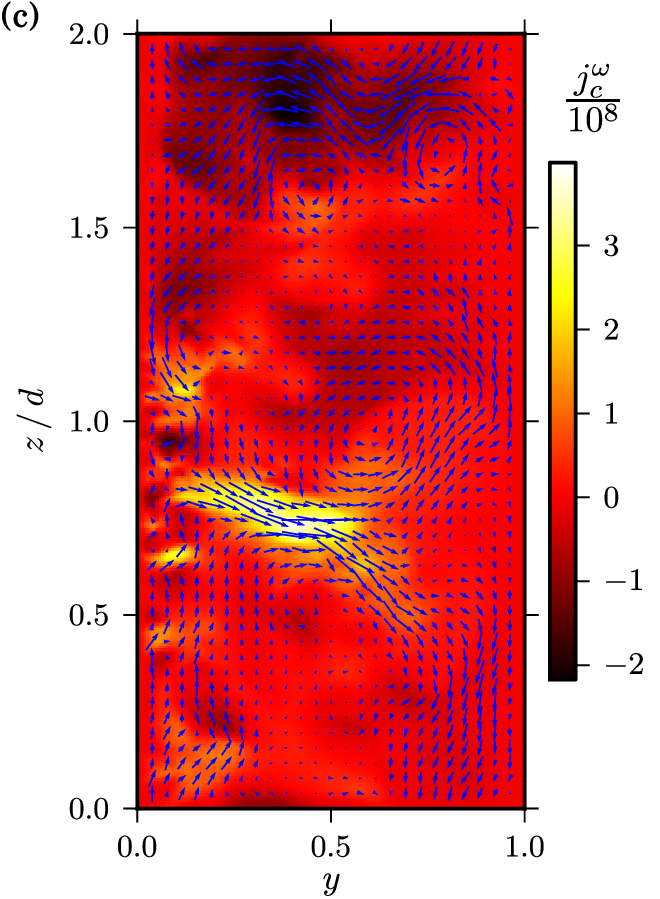}\includegraphics{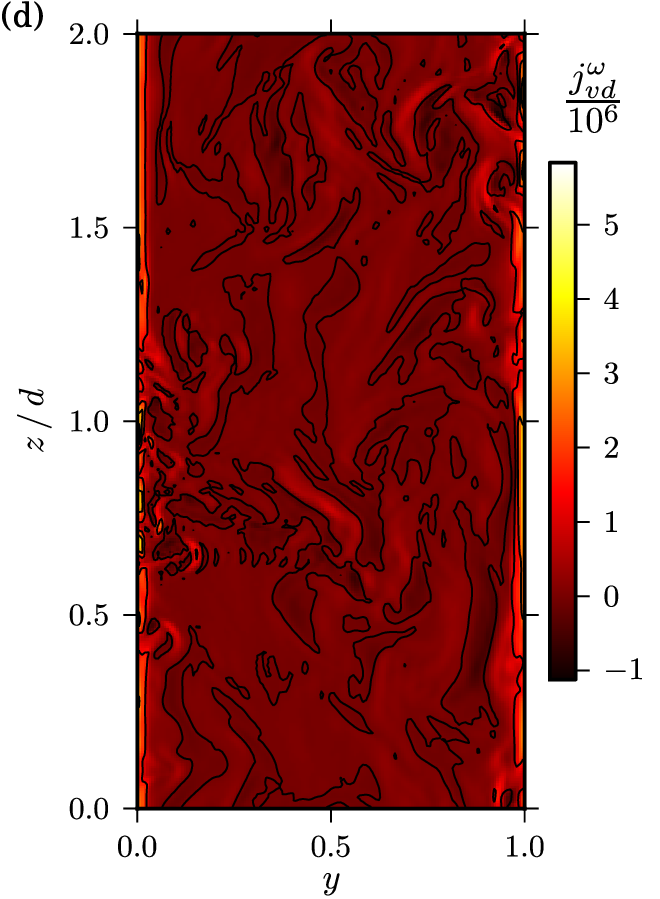}}
  \caption{Cross sections of local currents at a fixed angle for the flow situation depicted in figure \ref{fig:flow-viz}. Figure (a) shows the angular velocity current. (b), (c) and (d) show the terms that contribute to (a) according to table \ref{tab:jwcontrib}. Arrows indicate the in-plane velocity components. Note that the scales in (b) and (d) are about $10$ times and $100$ times, respectively, smaller than the one in (c). The local fluctuations exceed the mean $J^\omega=2.62\e{6}$ by two orders of magnitude.}
  \label{fig:contributions}
\end{figure}

We analyse the contributions of the different terms for the example of turbulent Taylor vortex flow with $\Rey_S=19737$ and stationary outer cylinder which was already shown in figure \ref{fig:current}. For this case the corresponding instantaneous fields are visualised in a $r$-$z$-plane for a fixed angle $\varphi$ in figure \ref{fig:flow-viz} and \ref{fig:contributions}. First the downstream velocity component $\omega=u_\varphi/r$ is shown with in-plane components represented by arrows (\mbox{figure \ref{fig:flow-viz}}), followed by a visualisation of the current $j^\omega$ (figure \ref{fig:contributions}(a)). Then the different contributions to the current are presented (figure \ref{fig:contributions}(b)-(d)). Using this flow case, some properties of the distribution of the contributions to $j^\omega$ can be discussed.

Even for the Reynolds number $\Rey_S=19737$, an inflow and outflow region is noticeable in the instantaneous flow visualisation (arrows in figure \ref{fig:contributions}(a)) indicating the presence of turbulent Taylor vortices. The flow shows a transition to the angular velocity prescribed by the cylinders in an extremely narrow region adjacent to the inner wall and a broader region near the outer wall, see figure \ref{fig:flow-viz}. Mushroom-like structures of angular velocity are pulled out from the wall, especially at the inner cylinder where they are smaller (cf. the magnifications in (b) and (c)). In a cylindrical plot, the mushrooms would correspond to azimuthally aligned high-$\omega$ streaks also reported by \cite{Dong2008a}. Within the scope of the $\omega$-current theory \citep{Eckhardt2007}, which is in analogy to thermal convection, these $\omega$-detachments from the boundaries correspond to thermal plumes in Rayleigh-B\'{e}nard flow.

The flattening of the angular velocity to a medium level in the bulk leads to a negative (positive) deviation from its laminar value in the inner (outer) region. Therefore, radial outflow implies a negative contribution $j^\omega_{rs}$ to the net current near the inner cylinder and a positive contribution near the outer cylinder. For the radial inflow, the signs are reversed, see figure \ref{fig:contributions}(b). In contrast, the consideration of the convective transport of $\Omega_b$ (not contributing to the mean $J^\omega$) leads to a current distribution that coincides with the radial flow, see figure \ref{fig:contributions}(c). Since $j^\omega_{rs}$ generally is much smaller than the convective transport of the laminar profile $j^\omega_c$, it only weakly influences the complete convective current $j^\omega_{rs}+j^\omega_c$ and the total current $j^\omega$, cf. figure \ref{fig:contributions}(a). Compared to $j^\omega_c$, the viscous derivative contribution to the current generally is about two orders of magnitude smaller and confined to the cylinder walls to such an extent that areas of maximal $j^\omega_{vd}$ are hard to identify in figure \ref{fig:contributions}(d). This indicates that strong radial $\omega$-gradients only occur in extremely narrow regions near the walls (cf. figure \ref{fig:flow-viz}). Most of the recognisable structures of $j^\omega_{vd}$ in the bulk are of negative sign, much weaker in amplitude than the ones expected at the walls, and negligible in relation to the convective current contributions. In summary, the spatial structures of the total current $j^\omega$ are  dominated by the convective contribution and thus by the occurrence of radial flow (weighted with $r^3\Omega_b$), see figure \ref{fig:contributions}(a). Consequently, the main part of the local current fluctuations (also in figure \ref{fig:current}(b)) originates from the transport term, which, however, does not contribute to the torque at all. These large contributions may also
explain the extreme local $j^\omega$ observed experimentally \citep{Huisman2012}. Fluctuations of the convective current contributing to the torque are much smaller, and only $\sim30$ times larger than the mean $J^\omega$ (figure \ref{fig:contributions}(b)).

\subsection{Spatial fluctuations}
\label{sec:spat-fluct}
The local values of $j^\omega$ give the local shear stress and 
their probability density function (PDF) near the wall can be compared to the PDF of the local wall shear stress $\tau_\omega$  
that was measured at the inner cylinder by \cite{Lathrop1992} and later at both cylinder walls by \cite{Lewis1999}. 
Both studies reported log-normal distribution functions in these cases. 

\begin{figure}
  \centerline{\includegraphics{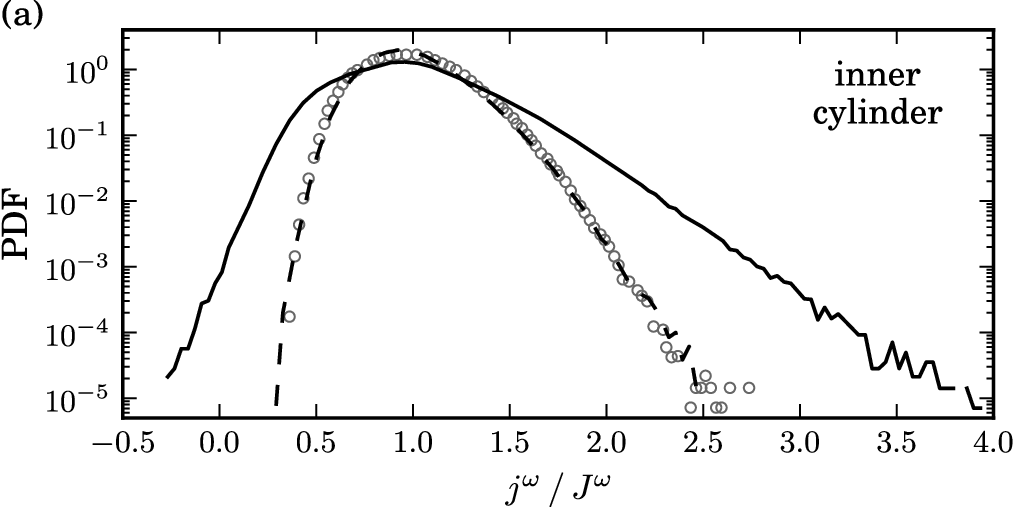}}
  \centerline{\includegraphics{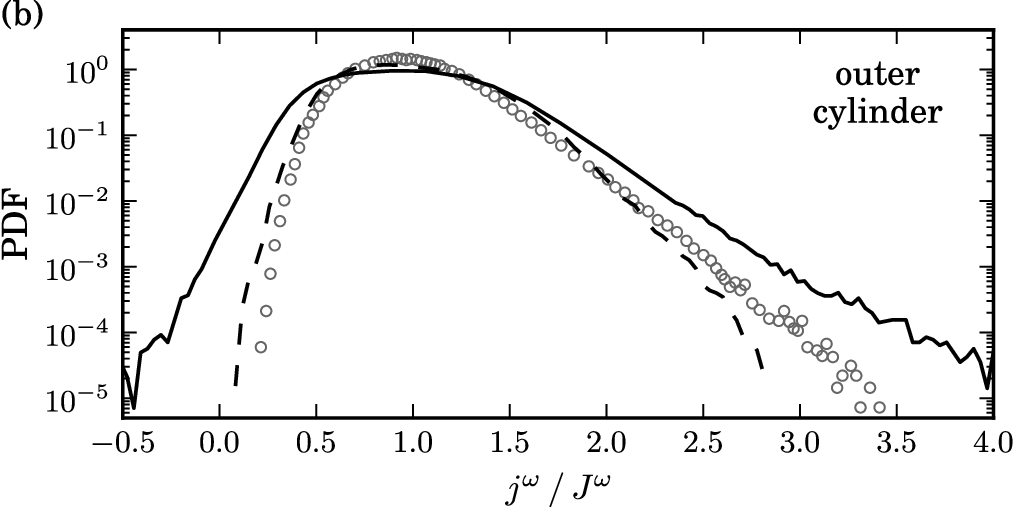}}
  \caption{Distribution of the spatial angular velocity current fluctuations obtained at the inner (a) and the outer cylinder (b). The circles correspond to measurements for $\Rey_S=81555$ \citep{Lewis1999} and the solid lines give numerical computations for $\Rey_S=29605$. The dashed lines indicate the same numerical data additionally averaged over $(n_\varphi\times n_z)=(7\times14)$ grid points.}
  \label{fig:fluct-comp}
\end{figure}

In order to compare to the experimental situation, the data for $\Rey_S=81555$ and a stationary outer cylinder from \citet[fig. 16]{Lewis1999} were digitised and included in Figure \ref{fig:fluct-comp}, where they are shown together with the corresponding PDFs (solid lines) computed for the highest Reynolds number achieved in the simulations ($\Rey_S=29605$) and for the outer cylinder at rest. 
Here, the number of evaluation points was $3/2$ times the number of modes in azimuthal and axial direction as needed for dealiasing \citep[p. 94]{Boyd2000}. One notes that experiment (open circles) and numerical data (continuous lines) do not agree, and that the
numerical data cover a wider range, including negative values. One also notes that while the fluctuations at inner and 
outer cylinder seem to be rather similar for the numerical simulations, they are different for the experimental data,
with the ones for the outer cylinder being wider.

\begin{figure}
  \centering
  \includegraphics{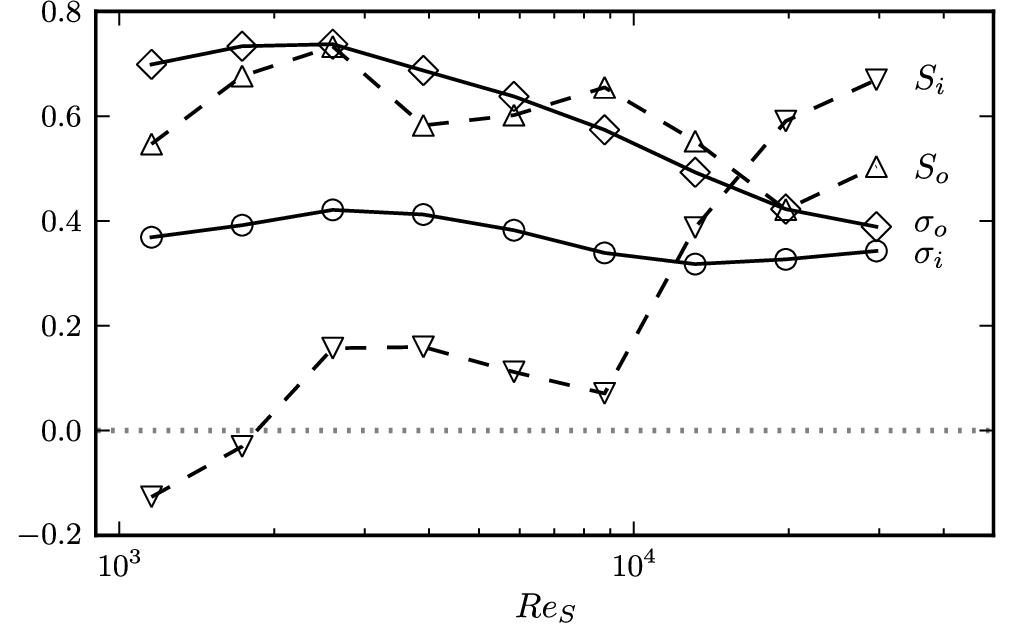}
  \caption{Standard deviation $\sigma_x$ (solid lines) and skewness $S_x$ (dashed lines) of $j^\omega$ fluctuations (rendered dimensionless with $J^\omega$) at the inner cylinder ($x=i$) and the outer cylinder ($x=o$) for different $\Rey_i$ and the outer cylinder at rest; here $\Rey_S=2/(1+\eta)\Rey_i$.}
 \label{fig:flc-scal}
\end{figure}

To determine whether the differences originate from a $\Rey_S$-dependence, we show in figure \ref{fig:flc-scal} the width (standard deviation $\sigma$) and the skewness $S$ of the $j^\omega$ fluctuations in relation to the mean $J^\omega$ for simulations at various $\Rey_S$. We find that the width of the distribution at the inner cylinder varies only slightly with $\Rey_S$. This agrees with the experimental observation that the standard deviation of the wall shear stress fluctuations at the inner cylinder is proportional to its mean \citep{Lathrop1992}. The width of $j^\omega$ fluctuations at the outer cylinder decreases slowly with $\Rey_S$ and approaches the inner cylinder level (figure \ref{fig:flc-scal}). For completeness also the skewness $S$ of the fluctuation distributions is given: While the outer cylinder skewness follows the trend of the corresponding standard deviation, the one for the inner cylinder starts with a negative value and increases with $\Rey_S$. In conclusion, the pronounced width difference between experimental and computed $j^\omega$ distributions (figure \ref{fig:fluct-comp}) is unlikely to come from the difference in Reynolds numbers.

Another difference between experiment and numerics is that in the numerical simulations we can obtain pointwise values
(within the numerical resolution) whereas in experiment there are spatial and temporal resolution issues to consider.
In order to explore the impact of a spatial averaging, we note that based on the dimensions of the hot film probe as given by
\citep{Lathrop1992} the experimental data are obtained with a spatial
resolution $(\delta_\varphi,\delta_z)$ in units of the gap width of $(4.9\e{-3},18\e{-3})$. In the numerical simulations
the grid point spacing gives a resolution of $(14\e{-3},4.6\e{-3})$. These resolutions must be evaluated relative to the observation that the axial correlation length of $j^\omega$ fluctuations is shorter than the azimuthal one, as is evident from the smaller axial characteristic size of turbulent structures in the $j^\omega$ visualisations in Figure \ref{fig:current}. Thus, a relatively small 
$\delta_z$ is especially important in order to resolve turbulent structures, and the experimentally achieved resolution would
already have to be improved at the Reynolds number of the simulations. Moreover, with an inner cylinder rotation of about $10\hz$ and provided a frequency response of the hot film probe of $10\khz$, cf. \cite{Lathrop1992}, the probe moves $17\e{-3}\,d$ during one data sample and therefore more than its width $\delta_\varphi=4.9\e{-3}$, which may lead to an additional azimuthal averaging. Finally, because of the three times larger $\Rey_S$, extreme events occurring in the experiment are more localised than the ones in the simulation and require a higher resolution for detection. These estimates suggest that the axial and possibly azimuthal averaging because of the relatively low resolution of the hot film probe leads to a cancellation of extreme $j^\omega$ events in the experimental distribution,
and hence to a narrower distribution. Moreover, because of the smaller length scales along the inner cylinder the 
effect should be more noticeable there, consistent with the asymmetry between inner and outer cylinder in the experimental data.

Since it is easy to introduce a spatial averaging in the numerical data, we computed $j^\omega$ fluctuations averaged over $7 \times 14$ grid points in the azimuthal $\times$ axial direction. The PDF's obtained this way are shown as dashed lines in 
Figure \ref{fig:fluct-comp} and reproduce the experimental data rather well, with almost perfect agreement for 
the inner cylinder and small deviations at the tails and the maximum for the outer cylinder. The poorer agreement for the latter distribution may originate from the stronger dependence of the standard deviation of $j^\omega/J^\omega$ on $\Rey_S$ for the outer cylinder (cf. figure \ref{fig:flc-scal}). Overall, averaging over plausible domains can account for a large part of the difference
between experiment and simulation.

\subsection{Temporal fluctuations}
\label{sec:temp-fluct}
\begin{figure}
  \centerline{\includegraphics{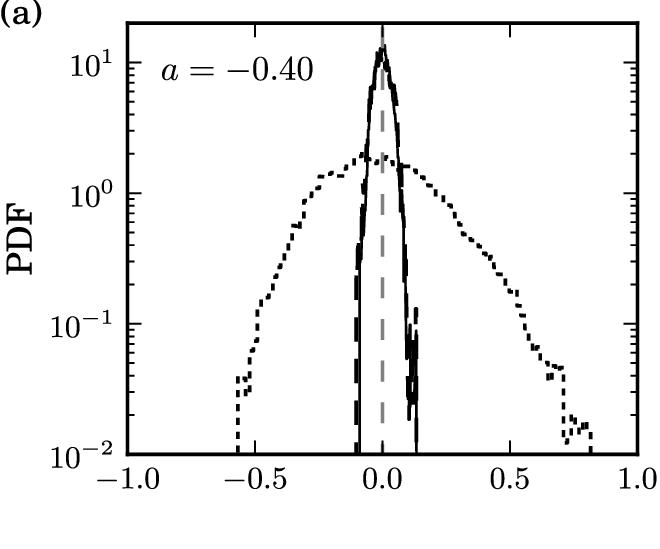}\includegraphics{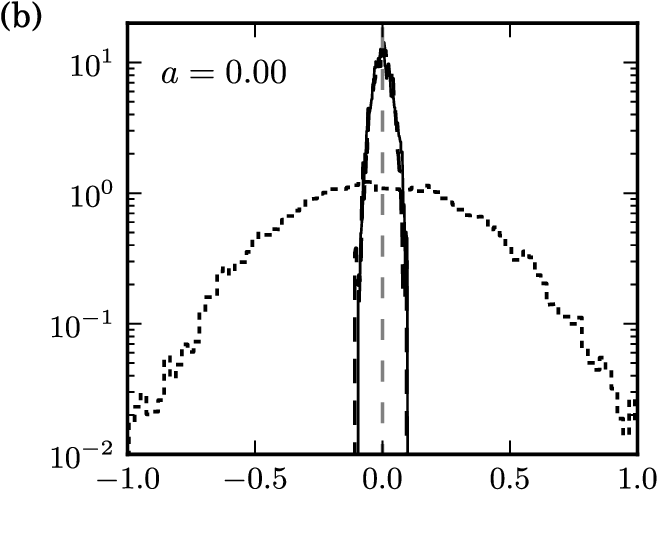}}
  \centerline{\includegraphics{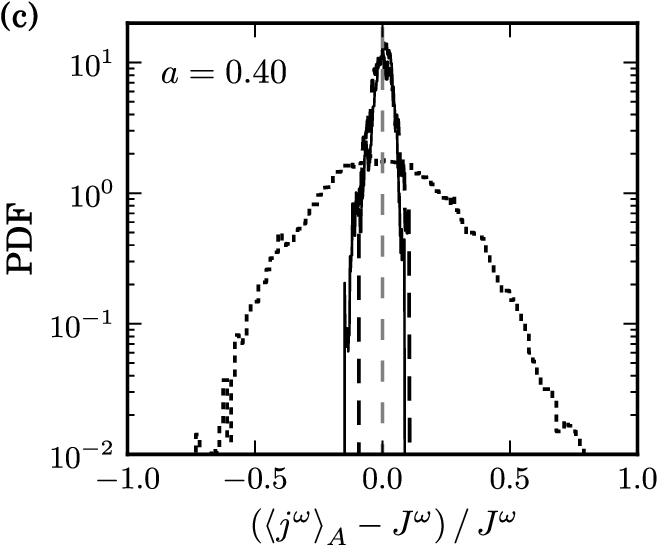}\includegraphics{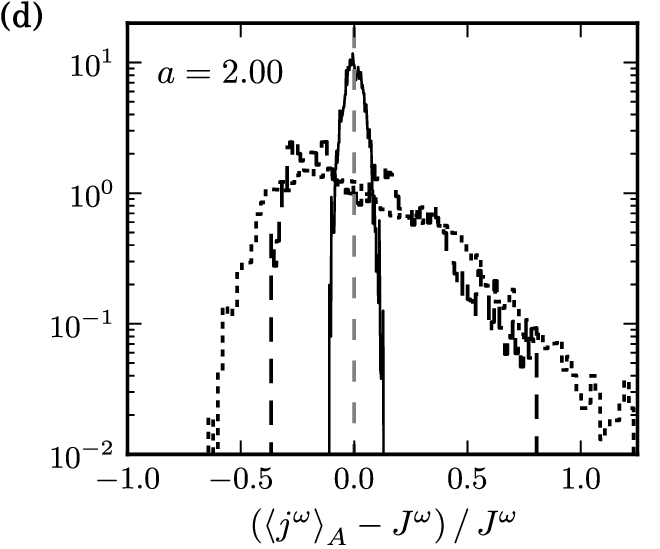}}
  \caption{Temporal fluctuations of the angular velocity current for $\Rey_S=19737$ and four rotation ratios $a$ computed at three radial locations: \protect\rule[.5ex]{.7cm}{0.5pt}, inner cylinder; $---$, outer cylinder; $\cdots$\,$\cdots$, midgap. Except for $a=2.00$ the PDFs of the fluctuations at the inner and outer cylinder lie almost on top of each other. Note that the scale in (d) is slightly increased.}
  \label{fig:temp-fluct}
\end{figure}

After analysing the spatial $j^\omega$ fluctuations we now turn to the temporal aspects of the current, after averaging over cylindrical surfaces $A(r)$. We focus on three locations, the inner cylinder, the outer cylinder and the middle of the gap, where the convective contribution in \eqref{eq:loc-current} dominates. We keep the shear Reynolds number $\Rey_S=19737$ fixed and vary the ratio of the
rotation rates, taking $a=-0.4$, $0.0$, $0.4$, and $2.0$ as representative examples. The PDF's for the fluctuations around the mean,
normalised by the mean, are shown in Figure \ref{fig:temp-fluct}.

As for the spatial $j^\omega$ distribution (figure \ref{fig:current}), the fluctuations are again significantly larger at midgap than near the inner or outer cylinder. For the first three values of $a$ the fluctuations at inner and outer cylinder agree and the
fluctuations are symmetric. For $a=2$ the distributions at inner and outer cylinder are different and the ones at midgap
and outer cylinder are strongly asymmetric. We also note that the fluctuations at the inner cylinder show little variation
with global variation. The fluctuations at midgap are strongest for outer cylinder at rest and are smaller for moderate
co- and counter-rotation. So what happens at $a=2$?

\begin{figure}
  \centerline{\includegraphics{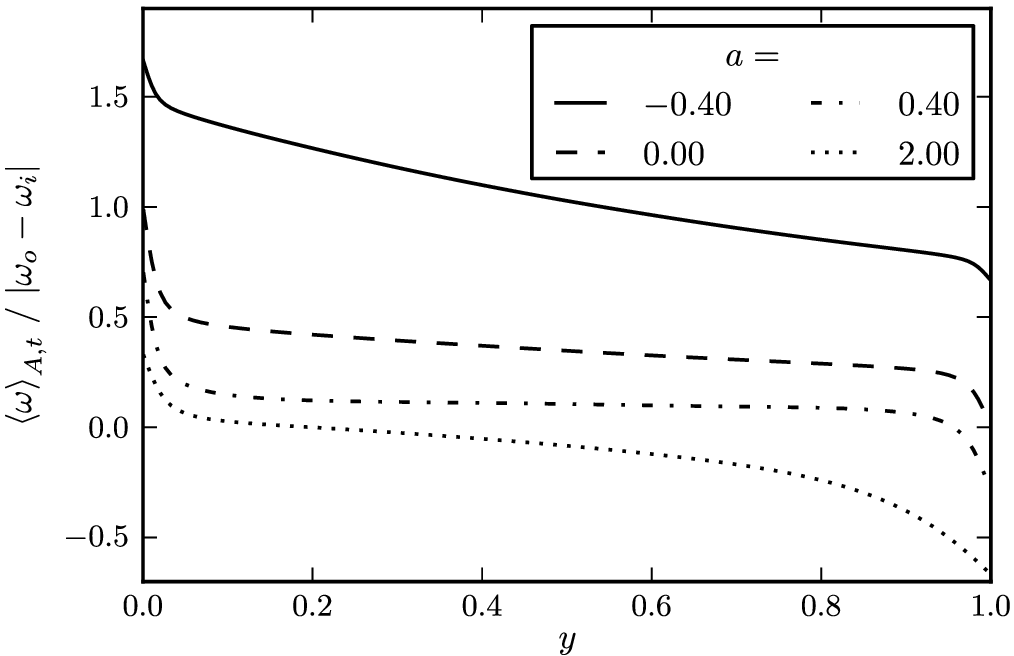}}
  \caption{Profiles of the mean angular velocity for the same shear $\Rey_S=19737$ and four different rotation ratios $a$ with $y=(r-r_i)/d$.}
  \label{fig:av-profiles}
\end{figure}

A first indication of a change in flow dynamics is provided by the radial profiles of angular velocity for the four rotation states discussed before, see figure \ref{fig:av-profiles}. Narrow gradient regions adjacent to both walls are clearly discernible for co- and moderate counter-rotation and stationary outer cylinder. In between the angular velocity exhibits a slight decline.  The small regions near the walls indicate thin boundary layers. In contrast, the angular velocity for strong counter-rotation exhibits a smooth transition from $\omega_o$ to its bulk level. One may suspect that this slow transition is due to a low turbulence level resulting in a thick boundary layer near the outer wall. This would suggest a connection to the known fact that for counter-rotation only a region near the inner cylinder is unstable according to Rayleigh's criterion \citep{Chandrasekhar1961}. The stable region adjacent to the outer cylinder increases with increasing counter-rotation. Although this stability analysis only holds for the laminar flow, one expects that the turbulence is less strongly driven near the outer cylinder. On the other hand, a laminar flow as such would be incompatible with the
fact that the current (\ref{eq:current}) has to be the same, independent of the radial position along which it is calculated. 
The strong fluctuations observed in the current at the outer cylinder (cf. figure \ref{fig:temp-fluct}(d)) are thus needed to 
compensate in the time average
the lower than average torque contribution in the less turbulent phase. These two flow states are then consistent with the bi-modal behaviour in the outer region recently observed by \citet{VanGils2011b} for $Re_S\sim10^6$. Similarly, at much lower Reynolds numbers, \citet{Coughlin1996} described oscillatory changes between laminar flow and turbulence in counter-rotating Taylor-Couette flow.

\section{Variations of torque with rotation rates}\label{sec:rot_depend}
Following the local characteristics we now turn to the variation of the global torque with rotation rates. More precisely, we study
how $Nu_\omega$, the torque expressed in units of its laminar value,  depends on the external driving, specifically
on the shear Reynolds number ($\Rey_S$) and on the global rotation state ($a=-\omega_o/\omega_i$). We will use the experimental
observation of \citep{VanGils2011,Paoletti2011} that the global variation can be represented by a scaling in $\Rey_S$ multiplied
by an amplitude that is a function of $a$ only.

\begin{table}
\centering
\begin{tabular}{ccccc}
$\Rey_S$ & $a$ & $(L,N,M)$ & $n_{sym}$ & $\tilde{\varepsilon}\Rey_S^{-2}Nu_\omega^{-1}$ \\
$1.155\e{3}$ & $-0.4$ to $2.0$ & $(38,46,32)$ & $1$ & $0.9949$ \\
$1.732\e{3}$ & $-0.4$ to $2.0$ & $(46,62,34)$ & $1$ & $0.9943$ \\
$2.599\e{3}$ & $-0.4$ to $2.0$ & $(46,78,34)$ & $1$ & $0.9932$ \\
$3.899\e{3}$ & $-0.4$ to $2.0$ & $(54,94,38)$ & $1$ & $0.9937$ \\
$5.848\e{3}$ & $-0.4$ to $2.0$ & $(62,110,44)$ & $1$ & $0.9945$ \\
$8.772\e{3}$ & $-0.4$ to $2.0$ & $(78,46,52)$ & $6$ & $0.9952$ \\
$1.316\e{4}$ & $-0.4$ to $2.0$ & $(78,46,56)$ & $6$  & $0.9949$ \\
$1.974\e{4}$ & $-0.4$ to $2.0$ & $(110,34,70)$ & $9$ & $0.9954$ \\
$2.961\e{4}$ & $-0.4$          & $(142,46,70)$ & $9$ & $1.0049$ \\
$2.961\e{4}$ & $-0.2$ to $1.0$ & $(142,38,70)$ & $9$ & $0.9942$ \\
$2.961\e{4}$ &           $2.0$ & $(126,46,70)$ & $9$ & $0.9982$ \\
\end{tabular}
\caption{Realised shear Reynolds numbers $\Rey_S$ with range of rotation ratio and employed resolution. For the outer cylinder at rest 
	 or alternatively the rotation ratio specified in the second column, the last column gives the 
	 quasi-Nusselt number computed from the dissipation rate in relation to $Nu_\omega$ which is suitable 
	 as convergence criterion.}
\label{tab:datrange}
\end{table}

For this purpose we realised $72$ simulations at nine shear Reynolds numbers ranging from $\Rey_S=1155$ to $\Rey_S=29605$ that are spaced by factors of $1.5$. The lower Reynolds number bound was chosen to ensure reasonably developed secondary flow, whereas the upper bound was dictated by computational restrictions. 
For each shear Reynolds number, eight different global rotation states with the focus on the regime of counter-rotation were selected. 
We include here two cases for co-rotation ($a=-0.4$ and $a=-0.2$), one with a stationary outer cylinder ($a=0$),
two for moderate counter-rotation ($a=0.2$ and $a=0.4$), one for equal but opposite rotation $\Rey_o=-Re_i$ ($a=\eta=0.71$) and 
two for high counter-rotation ($a=1.0$ and $a=2.0$).
For the highest $\Rey_S$ only one sixth or one ninth of the domain was simulated in the azimuthal direction (cf. table \ref{tab:datrange}). Calculations for a situation with stationary outer cylinder show that the differences
to a full simulation with $n_{sym}=1$ are below $1\%$, so that the reduction is justified. The resolution $(L,N,M)$ employed for each Reynolds number fulfils the exact computation of the global dissipation rate \eqref{eq:e_balance} within a relative error of $0.7\%$ at most, see the last column in table \ref{tab:datrange}.

\begin{figure}
  \centerline{\includegraphics{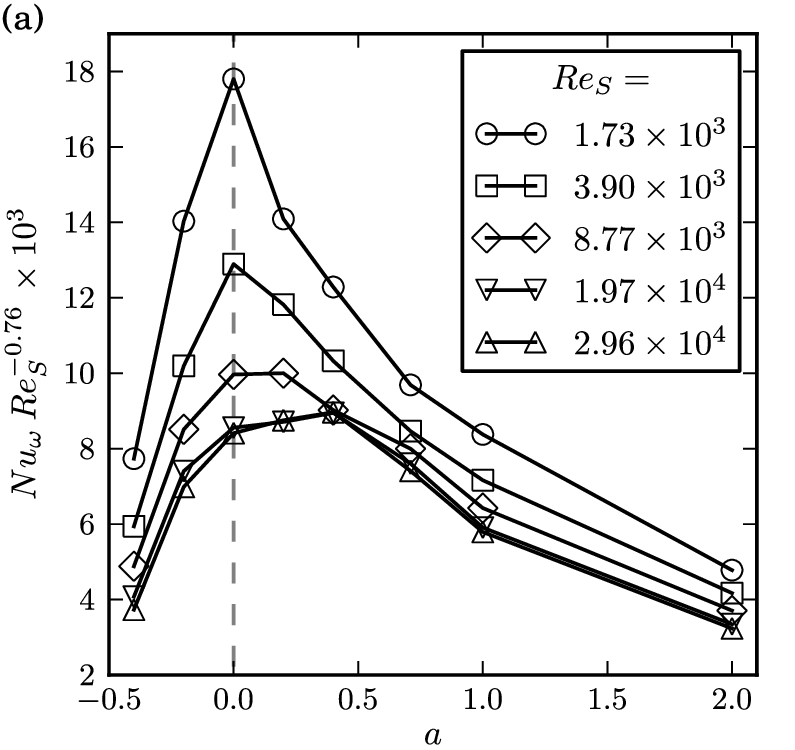}\includegraphics{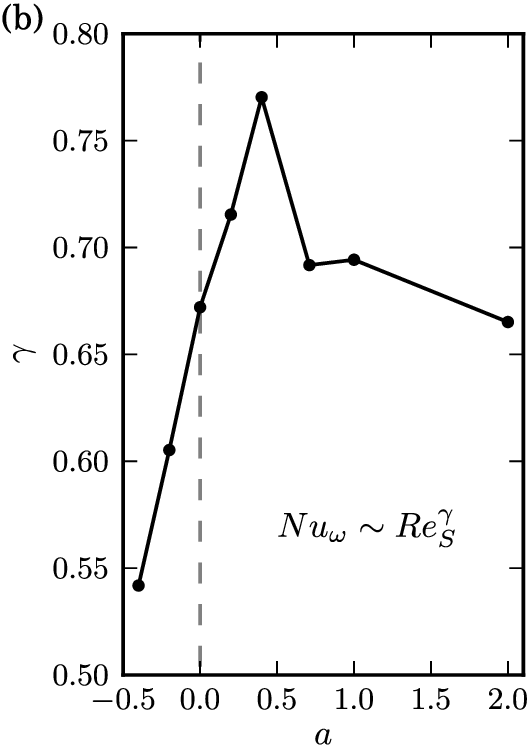}}
  \caption{Torque dependence on the shear and the global rotation: (a) Compensated plot of $Nu_\omega$ vs. rotation ratio $a$ for various $\Rey_S$. Flow states represented by symbols of the same shape feature the same shear. (b) Scaling exponents determined for each rotation ratio independently considering torques for $\Rey_S=1.32\e{4}$, $\Rey_S=1.97\e{4}$ and $\Rey_S=2.96\e{4}$.}
  \label{fig:nu-scal}
\end{figure}

For these flow cases the torques were computed at the inner cylinder and presented by means of $Nu_\omega$ compensated with $\Rey_S^{0.76}$ in figure \ref{fig:nu-scal}(a). The exponent $0.76$ is taken from the effective torque scaling $Nu_\omega \sim \Rey_S^{0.76}$ found by \cite{VanGils2011} for much higher Reynolds numbers between $3\e{5}$ and $2\e{6}$.
Each data point on a line in figure \ref{fig:nu-scal}(a) corresponds to a different global rotation state exhibiting the same shear. The changes with increasing $\Rey_S$ indicate that no global power law $Nu_\omega \sim \Rey_S^\gamma$ can be identified, as noted
previously by \citep{Lathrop1992, Lewis1999}. Nevertheless, the numerically determined effective torque scaling exponent, $\gamma=0.67$, for a stationary outer cylinder agrees well with $\gamma=0.69$ measured by \cite{Lewis1999} at $\Rey_S\approx 1.5\e{4}$. 
In addition to the variation with $Re_S$, the local scaling exponent $\gamma$ depends on the rotation ratio, see figure \ref{fig:nu-scal}(b). The exponent clearly decreases from moderate counter-rotation towards co-rotation, i.e. with decreasing rotation ratio. A similar trend is also discernible in effective exponents from high Reynolds number experiments \citep{VanGils2011b}. However, the authors note that both a trend in the exponents and a statistical scatter around an $a$-independent exponent are within the resolution of their experimental data. Presumably, the variations of $\gamma$ in \mbox{figure \ref{fig:nu-scal}(b)} are larger than in the experiment because of the relatively low $Re_S\sim10^4$, where the 
flow is not yet turbulence-dominated. Note, however, that the maximal scaling exponent, $\gamma=0.77$, in our simulations for $a=0.4$ already agrees well with the ultimate exponent, $\gamma=0.76$, measured by \citet{VanGils2011}.

Because of the complicated dependence of the torque on the shear and global rotation, a crossover in the rotation ratio that maximises
$Nu_\omega$ is observed: Initially, the transport contributing to the torque is most effective in a situation with stationary outer cylinder. However as the Reynolds number increases, the torque maximum becomes flatter and shifts towards moderate counter-rotation for $\Rey_S\geq2\e{4}$. In terms of the shear Reynolds number, this shift of the maximum towards $a\approx 0.4$ can be seen as a precursor for the torque maximisation at $a=0.368$ for $\eta=0.716$ \citep{VanGils2011b} and at $a=0.333$ for $\eta=0.7245$ \citep{Paoletti2011} as observed for even higher $\Rey_S$. However, both experimental studies did not report a shift of the torque maximum, presumably because the Reynolds number range they explore was much higher.

\begin{figure}
  \centerline{\includegraphics{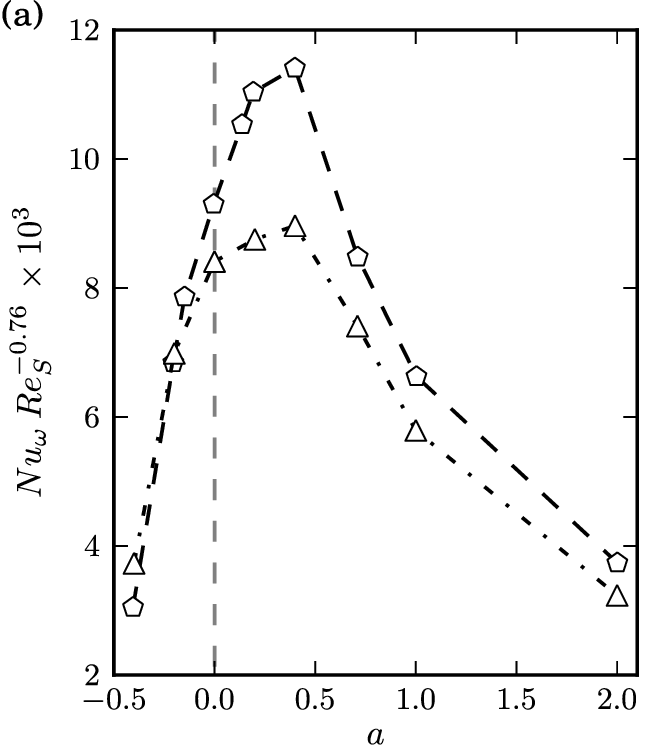}\includegraphics{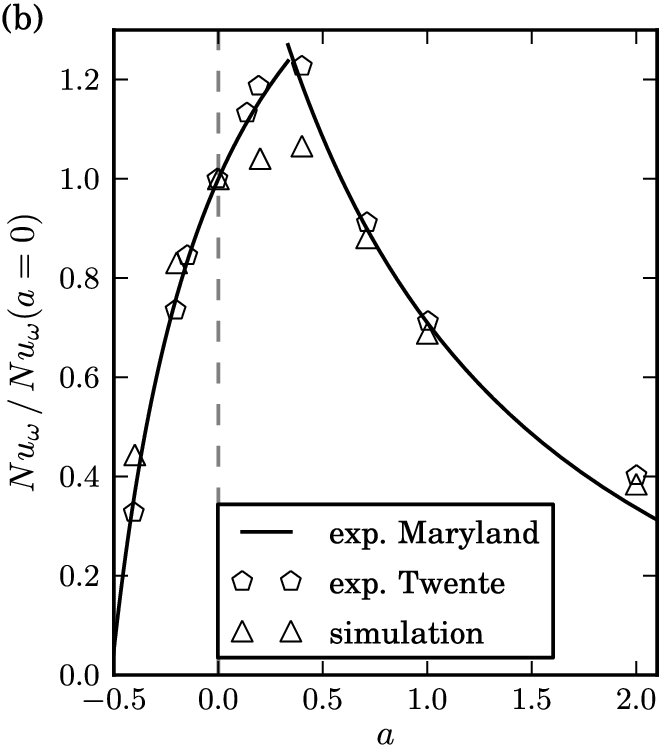}}
  \caption{Comparison of torque between the numerical simulation ($\eta=0.71$, $\Rey_S=2.96\e{4}$) and the experiments by \citet{VanGils2011} (Twente, $\eta=0.716$) and by \citet{Paoletti2011} (Maryland, $\eta=0.724$). (a) shows rescaled torques. Note that in the Twente experiment a rotation ratio-independent effective scaling $Nu_\omega \sim \Rey_S^{0.76}$ was found and the $a$-dependent prefactor of this scaling law is plotted here. (b) shows relative torques for both experiments and the simulation. The legend also applies to (a).}
  \label{fig:nu-comp}
\end{figure}

Observing the maximisation of the transport for approximately the same global rotation state, it is tempting to directly compare the functional dependence on the rotation ratio between the simulation at $\Rey_S=3\e{4}$ and the experiment even though they are separated by one to two orders of magnitude in Reynolds numbers. Appealing to the universal scaling found experimentally and the small change between the two highest $\Rey_S$ in figure \ref{fig:nu-scal}(a), it seems appropriate to compare the rescaled $Nu_\omega \Rey_S^{-0.76}$. For this type of comparison we assume that the mean dependence on $\Rey_S$ is divided out by the rescaling. 

In the experimentally found universal torque scaling \citep{VanGils2011}, only a rotation ratio-dependent proportionality factor in the power law absorbs the complete influence of the global rotation on the transport. This factor is given together with the corresponding calculated $Nu_\omega$ for the highest $\Rey_S$ in figure \ref{fig:nu-comp}(a). Qualitatively, the functional dependence on the rotation ratio agrees between simulation and experiment and the tails compare well even quantitatively. However, the torque maximum is not as pronounced in the simulation as in the experiment. This may be due to the relatively low Reynolds numbers accessible by simulations which is supported by the variation of the local scaling exponent in figure \ref{fig:nu-scal}(b). Its maximum for $a=0.4$ suggests that the torque maximum will become more pronounced in simulations for even higher $\Rey_S$.

A way to compare torque without the assumption of some scaling uses the torque of the inner cylinder at rest as a 
reference value.  \citet{Paoletti2011} observed that this torque ratio is almost independent of the shear Reynolds number and fitted the dependence on the global rotation by linear functions of $Ro^{-1}=\omega_o/(\omega_i-\omega_o)$ within four regions. Their functions for region $3$ and $4$ \citep[eq. (5),(6)]{Paoletti2011} are shown in figure \ref{fig:nu-comp}(b) together with data from the Twente experiment and from simulations. Except for the height of the $Nu_\omega$-maximum the computed torque values also agree with these empirical functions. Additionally, the observations of both experiments agree well with each other with a small deviation for strong counter-rotation. Note that this comparison only analyses the functional dependence on the global rotation based on relative torques. Therefore, possible differences in the absolute $Nu_\omega$ values can not be identified in figure \ref{fig:nu-comp}(b).

\section{Relation between torque and boundary-layer thicknesses}\label{sec:BLrelation}
Boundary layers play a key role in the analysis and description of global transport properties, and the numerical
simulations presented here provide a means to compare different definitions of boundary-layer thickness, and to pick the
one that is most suitable for the development of the theory. The observation that the torque shows no universal scaling with $Re_S$ in the investigated regime and, furthermore, depends on the mean rotation indicates that also the 
boundary-layer thickness
will have a more complicated dependence than an effective power law in $\Rey_S$ with a rotation-dependent prefactor, as observed experimentally for much higher $Re_S$ \citep{VanGils2011,Paoletti2011}. 

Thin layers near the cylinders, where the angular velocity changes from the wall to the bulk level, develop as a result of the applied shearing and are already visible in figure \ref{fig:av-profiles}. These angular velocity profiles illustrate boundary layers
(BL) due to azimuthal flow over the cylinder surface in the rotating frame with the corresponding cylinder at rest. Similarly, the axial flow over the cylinder surface is assumed to cause a second BL of an independent thickness \citep{Eckhardt2007}. Since the angular velocity profile directly enters the torque equation \eqref{eq:current}, only the first BL type will be analysed here.

\begin{figure}
  \centerline{\includegraphics{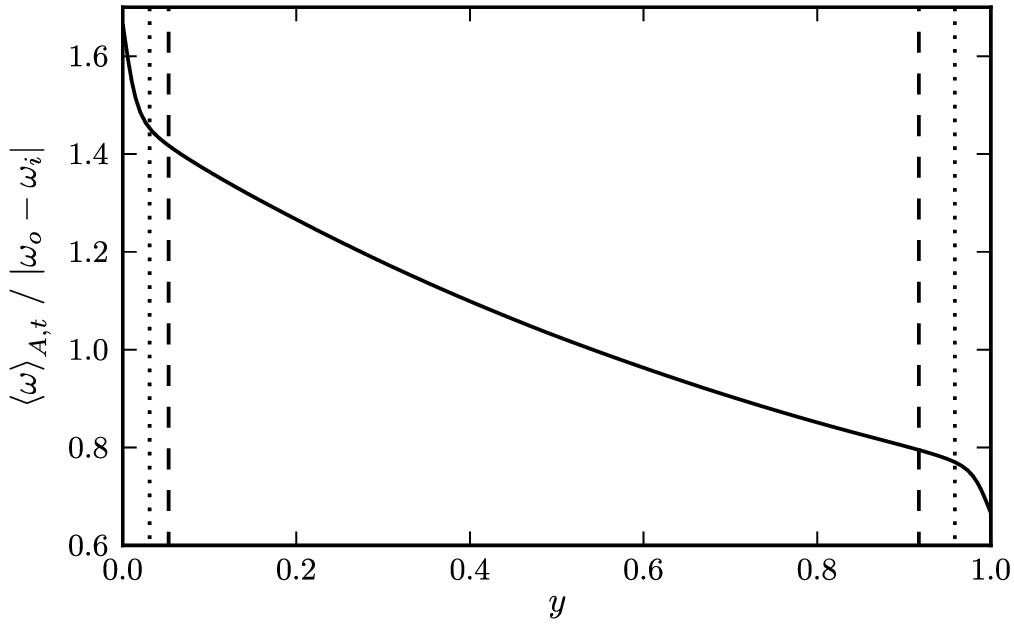}}
  \caption{Mean angular velocity profile for co-rotation ($a=-0.40$) and $Re_S=19737$ to exemplify the definition of boundary-layer thicknesses. The dashed lines indicate the slope thickness $\lambda_{sl}$, while the dotted lines indicate the effective boundary-layer thickness $\lambda_{vd}$ for the inner and outer cylinder BL, respectively.}
  \label{fig:BLprof}
\end{figure}

Common measures of the BL thickness for the flow over a flat plate include the $99\%$ velocity thickness and displacement thickness and require a free stream velocity $u_\infty$ to be computed \citep{Schlichting}. However, such an unambiguous free stream velocity does not exist in the TC flow because angular velocity profiles exhibit different slopes in the bulk depending on global rotation, cf. figure \ref{fig:av-profiles}. Another measure consists in the definition of a slope thickness based on the angular velocity, as proposed by \cite{Eckhardt2007}: For this purpose one relates the slope of the $\omega$-profile at the boundaries to a drop in angular velocity from the surface to a representative bulk value, $\left|\omega_x-\bar{\omega}\right|$. This leaves the boundary-layer thickness 
$\lambda_{sl}^{(x)}$ as an unknown,
\begin{equation}
  \partial_r \avg{\omega}_{A,t}\left|_{r_x}\right. = \frac{\left|\omega_x-\bar{\omega}\right|}{\lambda_{sl}^{(x)}}
  \; ; \qquad x=i, o \; .
  \label{eq:slBL}
\end{equation}
Here $\bar{\omega}$ denotes a bulk or centre level of the angular velocity as a point of reference. The result for the case 
of co-rotation ($a=-0.40$) with the definitions $\bar{\omega}=\avg{\omega(r_a)}_{A,t}$ and $r_a=(r_i+r_o)/2$ shows that this
overestimates the size of the region one would intuitively identify as BLs, see the dashed lines in figure \ref{fig:BLprof}. The 
obvious reason lies in the fact that the reference value $\bar{\omega}$ is estimated 
far from the boundary region, thereby giving a larger drop in angular velocity, and this has to be compensated by a
correspondingly larger value in $\lambda_{sl}$.

To overcome this problem, the definition of an ``effective boundary-layer thickness'' $\lambda_{vd}^{(x)}=\left|r_x-r_\lambda^{(x)}\right|$  with $x=i,o$ and radial locations, $r_\lambda^{(x)}$, corresponding to
\begin{equation}
  J^\omega_{vd}(r_\lambda^{(x)}) = \min_{r}\left\{J^\omega_{vd}\right\} + 
         0.1\,\left(J^\omega - \min_{r}\left\{J^\omega_{vd}\right\} \right) \; ,
  \label{eq:vdBL}
\end{equation}
can be employed \citep{Dong2008}. The subscript \textit{vd} indicates that it is based on the transition between viscous derivative $J^\omega_{vd}(r)=\avga{j^\omega_{vd}}$ and convective contribution to the current. It identifies the distance from the wall where the viscous derivative term exceeds its minimal value by $10\%$. Near this point the slope of the $\omega$-profile approaches its bulk value, but it does not necessarily vanish (cf. figure \ref{fig:BLprof}). The definition \eqref{eq:vdBL} is intrinsically close to the flow region description in the torque scaling theory by \cite{Eckhardt2007}. The measure $\lambda_{vd}$ corresponds to the BL regions one would expect from the shape of the $\omega$-profile for $a=-0.40$, see dotted lines in figure \ref{fig:BLprof}. Note that the relation $\lambda_{vd}<\lambda_{sl}$ for $a=-0.40$ is inverted for other rotation ratios.

So far, we introduced two different measures for angular velocity BL thicknesses which characterise the turbulent flow situation. Like the torque, these thicknesses vary with the externally applied shear and mean rotation. In the following, we analyse to which extent the variation of both BL thickness definitions agrees with the functional dependence of the torque. More precisely, we ask whether the knowledge of BL thicknesses suffices to predict the torque.

In order to study this issue, we employ the relation
\begin{equation}
  2 Nu_\omega = \sigma \frac{d}{\lambda^{(a)}} \; ,
  \label{eq:NuBL}
\end{equation}
with a weighted average $\omega$-BL thickness $\lambda^{(a)}$ and $\sigma=((1+\eta)/(2\sqrt{\eta}))^4$, which results from adding the two $J^\omega$ expressions \eqref{eq:current} at the boundaries and using equation \eqref{eq:slBL}, see \cite{Eckhardt2007}. 
Furthermore, one has to assume that the inner and outer boundary-layer thickness are related by
\begin{equation}
  \frac{r_i^3}{\lambda^{(i)}} \approx \frac{r_o^3}{\lambda^{(o)}} \approx \frac{r_a^3}{\lambda^{(a)}} \; ,
  \label{eq:BLrelat}
\end{equation}
which defines the average thickness $\lambda^{(a)}$. Relation \eqref{eq:BLrelat} originates from the observation that the constant current $J^\omega$ together with \eqref{eq:slBL} results in
\begin{equation}
 \frac{r_o^3}{\lambda^{(o)}} = \frac{|\omega_i-\bar{\omega}|}{|\omega_o-\bar{\omega}|} \, \frac{r_i^3}{\lambda^{(i)}} \; .
\end{equation}
In consequence, the assumption \eqref{eq:BLrelat} consists in the approximation that the ratio of angular velocity drops $|\omega_i-\bar{\omega}|/|\omega_o-\bar{\omega}|$ is close to unity. Here the weighted average BL thickness is defined for both BL thickness measures as 
\begin{equation}
  \lambda^{(a)} = \frac{r_a^3}{2} \left(\frac{\lambda^{(i)}}{r_i^3} + 
                          \frac{\lambda^{(o)}}{r_o^3} \right) \; ,
  \label{eq:wavg}
\end{equation}
consistent with relation \eqref{eq:BLrelat}. According to \eqref{eq:NuBL}, the product $2 Nu_\omega \lambda^{(a)}/d$ should be constant and close in value to $\sigma$. 

\begin{figure}
  \centerline{\includegraphics{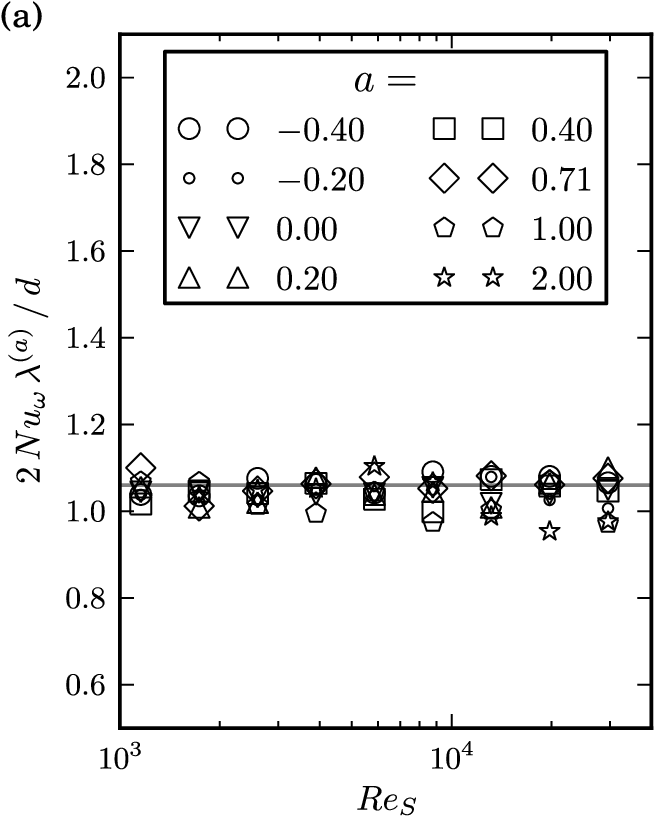}\includegraphics{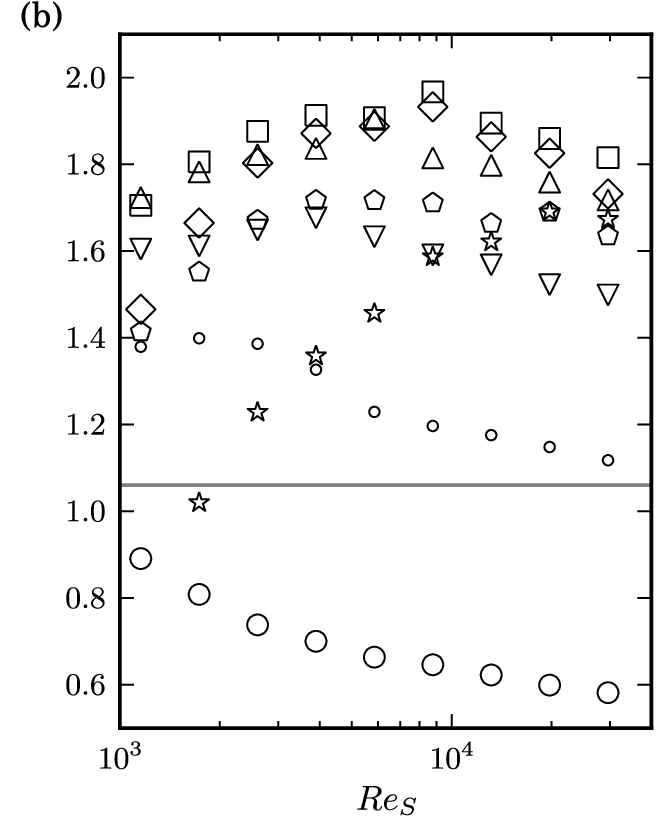}}
  \caption{Test of relation \eqref{eq:NuBL} between torque and average BL thickness $\lambda^{(a)}$ calculated from the slope BL thickness $\lambda_{sl}$ in (a) and from the effective BL thickness $\lambda_{vd}$ in (b). The grey line indicates the value $\sigma$ expected from theory. For symbols in (b) see legend in (a).}
  \label{fig:Nu-BLs}
\end{figure}

We first compute $\lambda^{(a)}$ using the slope thickness $\lambda_{sl}$ with a properly chosen bulk level of the angular velocity $\bar{\omega}=\avg{\omega(r_a)}_{A,t}$. We find that the deviation from expected relation \eqref{eq:NuBL} is not larger than $11\%$, see figure \ref{fig:Nu-BLs}(a). This high degree of conformance can be expected since the BL thickness measure $\lambda_{sl}$ was defined to be consistent with the description of the current $J^\omega$ and requires the knowledge of $\partial_r\avg{\omega}_{A,t}$ at the boundaries. The latter, however, can be directly used to calculate $Nu_\omega$. Nonetheless, the finding is not trivial because we needed the assumed relation \eqref{eq:BLrelat} which postulates that the ratio of angular velocity drops to the bulk level $\bar{\omega}$ equals unity.

The situation drastically changes when $Nu_\omega$ is instead related to a weighted average of the effective BL thickness $\lambda_{vd}$. The product $2 Nu_\omega \lambda^{(a)}/d$ shows a transient phase with increasing $Re_S$, especially for the cases of strong counter-rotation, see figure \ref{fig:Nu-BLs}(b). Then, the values vary less with $Re_S$, but deviate from $\sigma$ and differ for the various rotation ratios. A maximal proportionality coefficient is reached for moderate counter-rotation. The large deviations from $\sigma$ suggest that the BL thickness measure $\lambda_{vd}$ is not consistent with the characteristics of the current resulting in the torque.

In summary, the torque can be derived from BL thicknesses in the full parameter range considered here. However, the connection only works for a particular measure of the BL thickness, namely the slope thickness $\lambda_{sl}$ that is combined in a weighted average \eqref{eq:wavg}. Therefore, the knowledge of an angular velocity profile suffices to predict the torque. Although the definition of $\lambda_{vd}$ is based on the current $J^\omega$, it fails to explain the dependence of the torque on global rotation.

\section{Conclusions}\label{sec:conclusions}
Direct numerical simulations with accurate torque calculations were performed for Taylor-Couette flow and achieved Reynolds numbers $\Rey_S\sim3\e{4}$ not previously reported in literature. These simulations enabled comparisons to experimental investigations of turbulent Taylor-Couette flow at much higher Reynolds numbers. 

These direct numerical simulations show that Taylor-Couette turbulence in domains that carry only one pair of rolls and 
that cover only part of the circumference can give insights into the transition from the vortex dominated flows at lower
Reynolds numbers (up to about $\Rey_S\approx 10^4$) and the fully developed turbulence, which begins to show up near
$\Rey_S=3\e{4}$, the highest shear Reynolds numbers realised here. While the aspect ratio is rather important in the
Taylor-vortex regime, it seems to become less significant in the turbulent cases.

The statistical analysis of the local current $j^\omega$ showed that the fluctuations are wider than
experiments suggested and they reach to negative values. The reduction in width of the fluctuations could be explained
by an averaging over space: this suggests that future measures of these fluctuations require smaller probes. The fact that
the fluctuations can become negative indicates that locally in space and time there can be fluctuations that do not take energy
out of the mean flow but actually deposit energy there. This is consistent with observations on plane shear flow where a connection
to fluid structures could be established \citep{Schumacher2004}. 

The observation of the changes in mean profile with strong counter-rotation extend the observations of fluctuating
boundary layers from \cite{Coughlin1996} to higher Reynolds number. The onset of this phenomenon and the thickness of the
affected region require additional study that will be presented elsewhere.

We have also identified a definition of boundary-layer thickness that is compatible with the global torque
relation and that can be used in the corresponding analysis of the global transport relations \cite{Eckhardt2007}.
The definition is more complicated than the usual definitions because the reference value in the bulk is not a constant,
independent of the radius.
The variation of the mean profile in the middle
with radius is another remarkable observation, which, however, could also be gleaned from \citet{Wendt1933}. The origin of this 
variation, which is a difference between Rayleigh-B\'{e}nard and Taylor-Couette flow, remains unclear. 

The comparison to experiments was limited by a large gap in the Reynolds numbers: The simulations are
limited to $\Rey_S$ near several $10^4$, whereas experiments begin at several $10^5$. Nevertheless, several features
like the development of a torque maximum for counter-rotation could be studied.
Experimental investigations of the flow at moderate Reynolds numbers would be of great value to draw direct comparisons between experiments and simulations in the turbulent case. 
 
The torque maximum, the boundary-layer variations and the fluctuation statistics investigated here  
suggest that studies of turbulent Taylor-Coutte flow may  be just as rewarding as the analysis of the many bifurcations that lead from the laminar Couette flow to the turbulent flow state.

\section*{Acknowledgements}
We thank Siegfried Grossmann for stimulating discussions and Marc Avila for his help and support with the code.
Simulations of the highest Reynolds numbers were carried out at CSC Frankfurt.
This work was supported in part by Deutsche Forschungsgemeinschaft.

\bibliographystyle{jfm}

\bibliography{jfm-tc-paper}

\begin{thebibliography}{40}
\expandafter\ifx\csname natexlab\endcsname\relax\def\natexlab#1{#1}\fi

\bibitem[Andereck {\em et~al.\/}(1986)Andereck, Liu \& Swinney]{Andereck1986}
{\sc Andereck, C.~D., Liu, S.~S. \& Swinney, H.~L.} 1986 {Flow regimes in a
  circular Couette system with independently rotating cylinders}. {\em Journal
  of Fluid Mechanics\/} {\bf 164}, 155--183.

\bibitem[Bilson \& Bremhorst(2007)]{Bilson2007}
{\sc Bilson, M. \& Bremhorst, K.} 2007 {Direct numerical simulation of
  turbulent Taylor-Couette flow}. {\em Journal of Fluid Mechanics\/} {\bf 579},
  227--270.

\bibitem[Boyd(2000)]{Boyd2000}
{\sc Boyd, J.~P.} 2000 {\em {Chebyshev and Fourier Spectral Methods}\/}, 2nd
  edn. New York: Dover Pubns.

\bibitem[Burin {\em et~al.\/}(2010)Burin, Schartman \& Ji]{Burin2010}
{\sc Burin, M.~J., Schartman, E. \& Ji, H.} 2010 {Local measurements of
  turbulent angular momentum transport in circular Couette flow}. {\em
  Experiments in Fluids\/} {\bf 48}, 763--769.

\bibitem[Chandrasekhar(1961)]{Chandrasekhar1961}
{\sc Chandrasekhar, S.} 1961 {\em {Hydrodynamic and Hydromagnetic
  Stability}\/}, 1st edn. Oxford: Clarendon Press.

\bibitem[Chossat \& Iooss(1994)]{Chossat1994}
{\sc Chossat, P. \& Iooss, G.} 1994 {\em {The Couette-Taylor Problem}\/}. New
  York: Springer Verlag.

\bibitem[Coughlin \& Marcus(1996)]{Coughlin1996}
{\sc Coughlin, K. \& Marcus, P.~S.} 1996 {Turbulent Bursts in Couette-Taylor
  Flow}. {\em Physical Review Letters\/} {\bf 77}~(11), 2214--2217.

\bibitem[Dong(2007)]{Dong2007}
{\sc Dong, S.} 2007 {Direct numerical simulation of turbulent Taylor-Couette
  flow}. {\em Journal of Fluid Mechanics\/} {\bf 587}, 373--393.

\bibitem[Dong(2008{\natexlab{{\em a\/}}})]{Dong2008a}
{\sc Dong, S.} 2008{\natexlab{{\em a\/}}} {Herringbone streaks in
  Taylor-Couette turbulence}. {\em Physical Review E\/} {\bf 77}~(3), 035301.

\bibitem[Dong(2008{\natexlab{{\em b\/}}})]{Dong2008}
{\sc Dong, S.} 2008{\natexlab{{\em b\/}}} {Turbulent flow between
  counter-rotating concentric cylinders: a direct numerical simulation study}.
  {\em Journal of Fluid Mechanics\/} {\bf 615}, 371--399.

\bibitem[Dong(2009)]{Dong2009}
{\sc Dong, S.} 2009 {Evidence for internal structures of spiral turbulence}.
  {\em Phys. Rev. E\/} {\bf 80}, 067301.

\bibitem[Dong \& Zheng(2011)]{Dong2011}
{\sc Dong, S. \& Zheng, X.} 2011 {Direct numerical simulation of spiral
  turbulence}. {\em Journal of Fluid Mechanics\/} {\bf 668}, 150--173.

\bibitem[Dubrulle {\em et~al.\/}(2005)Dubrulle, Dauchot, Daviaud, Longaretti,
  Richard \& Zahn]{Dubrulle2005}
{\sc Dubrulle, B., Dauchot, O., Daviaud, F., Longaretti, P.-Y., Richard, D. \&
  Zahn, J.-P.} 2005 {Stability and turbulent transport in Taylor-Couette flow
  from analysis of experimental data}. {\em Physics of Fluids\/} {\bf 17}~(9),
  095103.

\bibitem[Eckhardt {\em et~al.\/}(2007{\natexlab{{\em a\/}}})Eckhardt, Grossmann
  \& Lohse]{Eckhardt2007a}
{\sc Eckhardt, B., Grossmann, S. \& Lohse, D.} 2007{\natexlab{{\em a\/}}}
  {Fluxes and energy dissipation in thermal convection and shear flows}. {\em
  Europhysics Letters (EPL)\/} {\bf 78}, 24001.

\bibitem[Eckhardt {\em et~al.\/}(2007{\natexlab{{\em b\/}}})Eckhardt, Grossmann
  \& Lohse]{Eckhardt2007}
{\sc Eckhardt, B., Grossmann, S. \& Lohse, D.} 2007{\natexlab{{\em b\/}}}
  {Torque scaling in turbulent Taylor-Couette flow between independently
  rotating cylinders}. {\em Journal of Fluid Mechanics\/} {\bf 581}, 221--250.

\bibitem[van Gils {\em et~al.\/}(2011{\natexlab{{\em a\/}}})van Gils, Huisman,
  Bruggert, Sun \& Lohse]{VanGils2011}
{\sc van Gils, D. P.~M., Huisman, S.~G., Bruggert, G.-W., Sun, C. \& Lohse, D.}
  2011{\natexlab{{\em a\/}}} {Torque Scaling in Turbulent Taylor-Couette Flow
  with Co- and Counterrotating Cylinders}. {\em Physical Review Letters\/} {\bf
  106}, 024502.

\bibitem[van Gils {\em et~al.\/}(2011{\natexlab{{\em b\/}}})van Gils, Huisman,
  Grossmann, Sun \& Lohse]{VanGils2011b}
{\sc van Gils, D. P.~M., Huisman, S.~G., Grossmann, S., Sun, C. \& Lohse, D.}
  2011{\natexlab{{\em b\/}}} {Optimal Taylor-Couette turbulence}. {\em Arxiv
  preprint\/} ~(arXiv:1111.6301v3).

\bibitem[Huisman {\em et~al.\/}(2012)Huisman, van Gils, Grossmann, Sun \&
  Lohse]{Huisman2012}
{\sc Huisman, S.~G., van Gils, D. P.~M., Grossmann, S., Sun, C. \& Lohse, D.}
  2012 {Ultimate Turbulent Taylor-Couette Flow}. {\em Physical Review
  Letters\/} {\bf 108}, 024501.

\bibitem[Jones(1985)]{Jones1985}
{\sc Jones, C.~A.} 1985 {The transition to wavy Taylor vortices}. {\em Journal
  of Fluid Mechanics\/} {\bf 157}, 135--162.

\bibitem[Koschmieder(1993)]{Koschmieder1993}
{\sc Koschmieder, E.~L.} 1993 {\em {B\'{e}nard Cells and Taylor Vortices}\/}.
  Cambridge University Press.

\bibitem[Lathrop {\em et~al.\/}(1992{\natexlab{{\em a\/}}})Lathrop, Fineberg \&
  Swinney]{Lathrop1992}
{\sc Lathrop, D.~P., Fineberg, J. \& Swinney, H.~L.} 1992{\natexlab{{\em a\/}}}
  {Transition to shear-driven turbulence in Couette-Taylor flow}. {\em Physical
  Review A\/} {\bf 46}~(10), 6390--6405.

\bibitem[Lathrop {\em et~al.\/}(1992{\natexlab{{\em b\/}}})Lathrop, Fineberg \&
  Swinney]{Lathrop1992a}
{\sc Lathrop, D.~P., Fineberg, J. \& Swinney, H.~L.} 1992{\natexlab{{\em b\/}}}
  {Turbulent Flow between Concentric Rotating Cylinders at Large Reynolds
  Number}. {\em Physical Review Letters\/} {\bf 68}~(10), 1515--1518.

\bibitem[Lewis \& Swinney(1999)]{Lewis1999}
{\sc Lewis, G.~S. \& Swinney, H.~L.} 1999 {Velocity structure functions,
  scaling, and transitions in high-Reynolds-number Couette-Taylor flow.} {\em
  Physical Review E\/} {\bf 59}~(5), 5457--67.

\bibitem[Marcus(1984)]{Marcus1984}
{\sc Marcus, P.~S.} 1984 {Simulation of Taylor-Couette flow. Part 1. Numerical
  methods and comparison with experiment}. {\em Journal of Fluid Mechanics\/}
  {\bf 146}, 45--64.

\bibitem[Meseguer(2003)]{Meseguer2003}
{\sc Meseguer, A.} 2003 {Linearized pipe flow to Reynolds number $10^7$}. {\em
  Journal of Computational Physics\/} {\bf 186}~(1), 178--197.

\bibitem[Meseguer {\em et~al.\/}(2007)Meseguer, Avila, Mellibovsky \&
  Marques]{Meseguer2007}
{\sc Meseguer, A., Avila, M., Mellibovsky, F. \& Marques, F.} 2007 {Solenoidal
  spectral formulations for the computation of secondary flows in cylindrical
  and annular geometries}. {\em The European Physical Journal - Special
  Topics\/} {\bf 146}, 249--259.

\bibitem[Meseguer {\em et~al.\/}(2009)Meseguer, Mellibovsky, Avila \&
  Marques]{Meseguer2009b}
{\sc Meseguer, A., Mellibovsky, F., Avila, M. \& Marques, F.} 2009 {Instability
  mechanisms and transition scenarios of spiral turbulence in Taylor-Couette
  flow}. {\em Physical Review E\/} {\bf 80}, 046315.

\bibitem[Moser {\em et~al.\/}(1983)Moser, Moin \& Leonard]{Moser1983}
{\sc Moser, R.~D., Moin, P. \& Leonard, A.} 1983 {A Spectral Numerical Method
  for the Navier-Stokes Equations with Applications to Taylor-Couette Flow}.
  {\em Journal of Computational Physics\/} {\bf 52}~(3), 524--544.

\bibitem[Paoletti \& Lathrop(2011)]{Paoletti2011}
{\sc Paoletti, M.~S. \& Lathrop, D.~P.} 2011 {Angular Momentum Transport in
  Turbulent Flow between Independently Rotating Cylinders}. {\em Physical
  Review Letters\/} {\bf 106}, 024501.

\bibitem[Pirr\`{o} \& Quadrio(2008)]{Pirro2008}
{\sc Pirr\`{o}, D. \& Quadrio, M.} 2008 {Direct numerical simulation of
  turbulent Taylor–Couette flow}. {\em European Journal of Mechanics
  B/Fluids\/} {\bf 27}, 552--566.

\bibitem[Racina \& Kind(2006)]{Racina2006}
{\sc Racina, A. \& Kind, M.} 2006 {Specific power input and local micromixing
  times in turbulent Taylor-Couette flow}. {\em Experiments in Fluids\/} {\bf
  41}, 513--522.

\bibitem[Ravelet {\em et~al.\/}(2010)Ravelet, Delfos \&
  Westerweel]{Ravelet2010}
{\sc Ravelet, F., Delfos, R. \& Westerweel, J.} 2010 {Influence of global
  rotation and Reynolds number on the large-scale features of a turbulent
  Taylor-Couette flow}. {\em Physics of Fluids\/} {\bf 22}~(5), 055103.

\bibitem[Recktenwald {\em et~al.\/}(1993)Recktenwald, L\"{u}cke \&
  M\"{u}ller]{Recktenwald1993}
{\sc Recktenwald, A., L\"{u}cke, M. \& M\"{u}ller, H.~W.} 1993 {Taylor vortex
  formation in axial through-flow: Linear and weakly nonlinear analysis}. {\em
  Physical Review E\/} {\bf 48}~(6), 4444--4454.

\bibitem[Riecke \& Paap(1986)]{Riecke1986}
{\sc Riecke, H. \& Paap, H.-G.} 1986 {Stability and wave-vector restriction of
  axisymmetric Taylor vortex flow}. {\em Physical Review A\/} {\bf 33}~(1),
  547--553.

\bibitem[Schlichting(1982)]{Schlichting}
{\sc Schlichting, H.} 1982 {\em {Grenzschicht-Theorie}\/}, 8th edn. Karlsruhe:
  Verlag G. Braun.

\bibitem[Schumacher \& Eckhardt(2004)]{Schumacher2004}
{\sc Schumacher, J. \& Eckhardt, B.} 2004 {Fluctuations of energy injection
  rate in a shear flow}. {\em Physica D\/} {\bf 187}, 370--376.

\bibitem[Stevens {\em et~al.\/}(2010)Stevens, Verzicco \& Lohse]{Stevens2010}
{\sc Stevens, R. J. A.~M., Verzicco, R. \& Lohse, D.} 2010 {Radial boundary
  layer structure and Nusselt number in Rayleigh-B\'{e}nard convection}. {\em
  Journal of Fluid Mechanics\/} {\bf 643}, 495--507.

\bibitem[Taylor(1923)]{Taylor1923}
{\sc Taylor, G.~I.} 1923 {Stability of a Viscous Liquid contained between Two
  Rotating Cylinders}. {\em Philosophical Transactions of the Royal Society of
  London. Series A\/} {\bf 223}, 289--343.

\bibitem[Taylor(1936)]{Taylor1936}
{\sc Taylor, G.~I.} 1936 {Fluid Friction between Rotating Cylinders. I. Torque
  Measurements}. {\em Proceedings of the Royal Society of London. Series A\/}
  {\bf 157}~(892), 546--564.

\bibitem[Wendt(1933)]{Wendt1933}
{\sc Wendt, F.} 1933 {Turbulente Str\"{o}mungen zwischen zwei rotierenden
  konaxialen Zylindern}. {\em Ingenieur-Archiv\/} {\bf 4}, 577--595.

\end{thebibliography}

\end{document}